\documentclass[sigconf,nonacm]{acmart}

\usepackage{appendix}
\usepackage{amsmath}

\usepackage{amssymb,amsfonts}
\usepackage{algorithmic}
\usepackage{graphicx}
\usepackage{textcomp}
\usepackage{standalone}
\usepackage{xcolor}
\def\BibTeX{{\rm B\kern-.05em{\sc i\kern-.025em b}\kern-.08emT\kern-.1667em\lower.7ex\hbox{E}\kern-.125emX}}
\usepackage[utf8]{inputenc}
\usepackage{booktabs} % For formal tables
\usepackage{graphicx}
\usepackage{units}
\usepackage{blindtext}
\usepackage{multirow}
\usepackage{xcolor,xspace}
\usepackage{amsmath}
\usepackage{paralist}
\usepackage{mdframed}
\usepackage{pict2e}
\usepackage{listings}
\usepackage{color}
\usepackage{threeparttable}
\usepackage{array}
\usepackage{booktabs}
\usepackage{pifont}
\usepackage{url}
\usepackage{comment}
\usepackage{amsmath}
\usepackage{lipsum}
\usepackage{arydshln}
\usepackage{flushend}
\usepackage{latexsym}
\usepackage{enumitem}
\usepackage{fancyhdr}
\usepackage[english]{babel}
\usepackage{tabularx}
\input{glyphtounicode}

\usepackage[ruled]{algorithm2e}
\usepackage{algorithmic}
\usepackage{pgfplots,subcaption}
\pgfplotsset{compat=newest}
\usetikzlibrary{patterns}

\usepackage{url}
\usepackage[
    n,
    advantage,
    operators,
    sets,
    adversary,
    landau,
    probability,
    notions,
    logic,
    ff,
    mm,
    primitives,
    events,
    complexity,
    asymptotics,
    keys]{cryptocode}

\usepackage{tikz}                    
\usetikzlibrary{arrows, patterns, automata, positioning}
\newcommand{\ignore}[1]{}

\newcommand{\edit}[1]{\textcolor{black}{#1}}

\newcommand{\prv}{{\ensuremath{\sf{\mathcal Prv}}}\xspace}
\newcommand{\vrf}{{\ensuremath{\sf{\mathcal Vrf}}}\xspace}

\newcommand{\RA}{{\textit{RA}}\xspace}

\newcommand{\CFA}{{\textit{CFA}}\xspace}

\newcommand{\DFA}{{\textit{DFA}}\xspace}
\newcommand{\acron}{\textit{{SABRE}}\xspace}

\renewcommand\adv{\ensuremath{\sf{\mathcal Adv}}\xspace}

\newcommand{\cflog}{\ensuremath{\mathcal{C}F_{Log}}\xspace}
\newcommand{\slice}{\ensuremath{\mathcal{C}F_{Slice}}\xspace}
\newcommand{\cflognospace}{\ensuremath{\mathcal{C}F_{Log}}}

\newcommand{\ctrldata}{$addr_{target}$\xspace}

\newcommand{\appbin}{\textit{App.elf}\xspace}

\newlist{myenumerate}{enumerate}{1}
\setlist[myenumerate]{
  label=(\arabic*),
  align=left,
  leftmargin=*,
  nosep,
}

\newlist{myitemize}{itemize}{1}
\setlist[myitemize]{
    label=$\bullet$,
    align=left,
    leftmargin=*,
    nosep,
}

\definecolor{mGreen}{rgb}{0,0.6,0}
\definecolor{mGray}{rgb}{0.5,0.5,0.5}
\definecolor{mPurple}{rgb}{0.58,0,0.82}
\definecolor{backgroundColour}{rgb}{0.95,0.95,0.92}

\lstdefinestyle{CStyle}{
    backgroundcolor=\color{white},   
    commentstyle=\color{mGreen},
    keywordstyle=\color{blue},
    numberstyle=\tiny\color{black},
    stringstyle=\color{blue},
    basicstyle=\ttfamily\footnotesize,
    breakatwhitespace=false,         
    breaklines=true,                 
    captionpos=b,                    
    keepspaces=true,                 
    numbers=left,                    
    numbersep=5pt,      
    frame=single,
    xleftmargin=1.75em,
    xrightmargin=1.75em,
    framexleftmargin=1em,
    captionpos=b,
    showspaces=false,                
    showstringspaces=false,
    showtabs=false,                  
    tabsize=2,
    float=tp,
    language=C
}

\begin{document}

\title{On the Verification of Control Flow Attestation Evidence}% via Binary Analysis} %add context of CFA to title

% \author{Anonymous Authors}

\author{Adam Ilyas Caulfield}
\affiliation{%
\institution{University of Waterloo}
\city{Waterloo}
\country{Canada}
}
\email{acaulfield@uwaterloo.ca}

\author{Norrathep Rattanavipanon}
\affiliation{%
\institution{Prince of Songkla University}
\city{Phuket}
\country{Thailand}
}
\email{norrathep.r@phuket.psu.ac.th}

\author{Ivan De Oliveira Nunes}
\affiliation{%
\institution{University of Zurich}
\city{Zurich}
\country{Switzerland}
}
\email{ivan.deoliveiranunes@uzh.ch}

\renewcommand{\shortauthors}{Adam Ilyas Caulfield, Norrathep Rattanavipanon, and Ivan De Oliveira Nunes}

% \author{Adam Caulfield$^{1}$ \and Norrathep Rattanavipanon$^{2}$ \and Ivan De Oliveira Nunes$^{1}$}

% \institute{$^{1}$Rochester Institute of Technology, USA \\ $^{2}$Prince of Songkla University, Phuket Campus, Thailand}

%\authorrunning{Adam Caulfield et al.}
%\titlerunning{Run-time Attestation and Auditing: The Verifier's Perspective}

% \pagestyle{plain}

%%\vspace{-2em}
\begin{abstract}
%%% slightly shortened
In run-time attestation schemes, including Control Flow Attestation (\CFA) and Data Flow Attestation (\DFA), a remote Verifier (\vrf) requests a potentially compromised Prover device (\prv) to generate evidence of its execution control flow path (in \CFA) and optionally execution data inputs (in \DFA). Recent advances in this space also guarantee that \vrf eventually receives run-time evidence from \prv, even when \prv is fully compromised. Reliable delivery, in theory, enables \textit{run-time auditing} in addition to attestation, allowing \vrf to examine run-time compromise traces to pinpoint/remediate attack root causes. However, \vrf's perspective in this security service remains unexplored, with most prior work focusing on the secure generation of authentic run-time evidence on \prv.

In this work, we argue that run-time attestation/auditing is only effective if \vrf can analyze the received evidence. From this premise, we characterize different types of evidence produced by run-time attestation/auditing architectures based on \vrf's ability to use them for vulnerability detection/remediation.
As a case study, we propose \acron: a \underline{S}ecurity \underline{A}nalysis and \underline{B}inary \underline{R}epair \underline{E}ngine that enables \vrf to use run-time evidence to detect control flow attacks, to pinpoint specific instructions that corrupted control data, and to automatically generate binary patches to buffer overflow and use-after-free vulnerabilities without source code knowledge.~\footnote{This work appeared in The 18$^{\texttt{th}}$ ACM Conference on Security and Privacy in Wireless and Mobile Networks (ACM WiSec 2025) with the following title: \textit{Run-time Attestation and Auditing: The Verifier’s Perspective}.}

\end{abstract}

%    <concept>
%        <concept_id>10002978.10003006</concept_id>
%        <concept_desc>Security and privacy~Systems security</concept_desc>
%        <concept_significance>500</concept_significance>
%        </concept>
%  </ccs2012>

% \ccsdesc[500]{Security and privacy~Systems security}

% \keywords{Embedded Systems, Control Flow Attestation, Run-time Auditing}

% \begin{IEEEkeywords}
% Control Flow Attestation, Systems Security, Embedded Systems
% \end{IEEEkeywords}

\copyrightyear{2025}
\acmYear{2025}
\setcopyright{acmlicensed}\acmConference[WiSec 2025]{18th ACM Conference on Security and Privacy in Wireless and Mobile Networks}{June 30-July 3, 2025}{Arlington, VA, USA}
\acmBooktitle{18th ACM Conference on Security and Privacy in Wireless and Mobile Networks (WiSec 2025), June 30-July 3, 2025, Arlington, VA, USA}
\acmDOI{10.1145/3734477.3734710}
\acmISBN{979-8-4007-1530-3/2025/06}

\maketitle

\section{Introduction}
Embedded devices exist at the edge of larger systems, often serving as remotely operated sensors and on-demand actuators. Due to low energy and cost requirements, they are commonly implemented using low-end micro-controller units (MCUs) that lack advanced architectural security features (e.g., virtual memory or memory management units). Consequently, MCUs are more vulnerable and frequently targeted by cyber attacks~\cite{nafees2023smart,kayan2022cybersecurity}.

Given their low budget for preventative security features, MCUs need inexpensive means to prove their software integrity to back-ends that rely on their services (e.g., a device owner or control center). Towards this goal, Remote Attestation (\RA)~\cite{coker2011principles} has been proposed to give a resource-rich back-end -- called a Verifier (\vrf) -- the ability to remotely assess the software state of a resource-constrained Prover MCU (\prv). In \RA, \vrf challenges \prv to return a cryptographic proof of \prv's currently installed software binary. Only after receiving an authentic proof will \vrf trust that \prv is installed with the correct software. 

Although \RA can effectively detect illegal code modifications, it cannot detect run-time attacks that do not modify code~\cite{cflat}. For example, an adversary (\adv) could exploit a memory vulnerability (e.g., a buffer overflow~\cite{cowan2000buffer}) to corrupt \textit{control} data, including return addresses, indirect jump targets, and function pointers. Consequently, \adv can chain out-of-order instruction sub-sequences to execute (often Turing-complete) malicious behavior~\cite{jop,rop}.
Since these attacks do not modify code, they are oblivious to \RA.  

To address this limitation, \RA can be extended into Control Flow Attestation (\CFA).
In \CFA, \vrf obtains cryptographic proof of both the installed software image and the exact control flow path followed in the most recent execution of the software~\cite{sok_cfa_cfi}.
This evidence and the respective proof are generated/authenticated by a root of trust (RoT) within \prv that securely records control flow transfer destinations.
Therefore, \CFA evidence can, in principle, allow \vrf to detect both code modification and control-flow hijacking attacks.
Recent architectures augment \CFA to obtain \textit{run-time auditing}~\cite{acfa,traces} by guaranteeing the delivery of \CFA evidence to \vrf even when \prv's software is compromised. See Sec.~\ref{sec:bg} for details on \CFA and run-time auditing RoTs.

We observe that prior work in this area~\cite{sok_cfa_cfi} has focused on secure design/implementation of \prv RoTs, often overlooking \vrf's role.
%Current efforts focusing on \vrf's role look into privacy-preserving path verification via Zero-Knowledge proofs~\cite{debes2023zekra} or addressing scalability concerns of verification~\cite{rage}.
%
%In a similar vein, \prv-focused prior work proposes various \CFA evidence formats that allow \vrf to compute a boolean verification result of the reported path. 
To our knowledge, no systematic analysis has explored \vrf's ability to interpret \CFA evidence beyond basic path validation (i.e., determining whether the reported control-flow path is valid). 
However, in practice, after deeming a path invalid (thus \prv compromise), \vrf would need to use \CFA evidence to (1) identify the root cause of this compromise and (2) implement patches to mitigate it. Neither of these has been thoroughly investigated in the literature.
The challenge increases when \prv's software contains proprietary dependencies/third-party libraries with unavailable source code: a common occurrence in embedded system software chains.
%%% Adam: based on the USENIX Reviewer 1, should we cite something here?

Given the prevalence of memory-unsafe languages in embedded software and their various associated exploits~\cite{cisa-urgent-need-for-memory-safety}, we argue that automated methods to identify unknown memory safety vulnerabilities from attested run-time evidence can result in timely detection of otherwise oblivious attacks and elimination of their root causes.
%As \vrf's perspective has not been a focus in prior work, this work further illustrates that prior \CFA RoTs generate evidence in different formats (e.g., {\it verbatim} control flow path vs. a hash chain of all control flow transfers and hybrids thereof), some of which are insufficient to support the identification and patching of certain exploits.
%Furthermore, 
While techniques have been proposed for root cause vulnerability analysis in high-end devices (e.g., based on Intel PT traces~\cite{intel_pt}), 
none of them can be used to analyze execution traces produced by \CFA.
%there are no frameworks for analysis of execution traces produced by \CFA in MCUs. 
In particular, these techniques~\cite{yagemann2021arcus,yagemann2021automated,zeng2022palantir,liu2022seeker} require initial memory snapshots for a remote server to perform a root cause analysis.
This requirement is unlikely to be satisfied in the \CFA settings for two reasons.
First, this would require \prv to transmit potentially large memory snapshots during \CFA, conflicting with the resource-constrained nature of MCUs~\cite{sok_cfa_cfi}.
% OAK: is below true? I thought CFA can also be a challenge-response protocol.
%% adam: removing this because i'm not sure if its a good argument anymore...
% Second, since \CFA makes no assumption about when a request occurs, \prv would need to store and retain the snapshot until a \CFA request arrives. 
Second, this snapshot must be stored by \prv during the attested execution. This, in turn, unnecessarily reduces memory available to regular tasks on \prv. We expand on these points in Sec.~\ref{sec:rw}.
Due to these conflicts, existing root cause analysis frameworks are not directly compatible with \CFA.
% OAK: I dont understand what "no augmentations" mean so I remove it for now but feel free to bring it back if its elaborated more.
%and requires no augmentations to \prv's RoT or evidence.}
%
Our work bridges these gaps with the following contributions:
%
%this work presents \acron: a \underline{D}etection, \underline{A}nalysis, and \underline{B}inary \underline{R}epair \underline{E}ngine designed for MCUs in Control Flow Attestation/Auditing. \acron provides automated processes enabling a \vrf to conduct further analysis on the binary executable running on a \prv MCU, along with the run-time evidence obtained from a \CFA RoT on \prv. After detecting a control flow attack from the run-time evidence, \acron aims to locate and patch the buffer overflow vulnerabilities within the binary that are responsible. Through the processes available in \acron, \vrf can do the following: detect a malicious program path present in the run-time evidence, identify the corrupted control data responsible for deviating from a valid path, symbolically execute a program slice to identify the exact memory instruction utilized during the buffer overflow, automatically generate a patch for the compromised memory instruction, and validate the patch effectiveness against attempts to corrupt the same control data. 
%
%The contributions of this work are summarized as follows:
\begin{myenumerate}
    \item We analyze and classify the types of run-time evidence generated by existing \CFA RoTs. Their trade-offs are examined in detail with particular focus on their ability to support \vrf's remote vulnerability detection and remediation.
    \item As a case study, we present \acron: a \underline{S}ecurity \underline{A}nalysis and \underline{B}inary \underline{R}epair \underline{E}ngine based on \CFA evidence. \acron combines binary analysis with \CFA run-time evidence to automatically locate and patch buffer overflow and use-after-free vulnerabilities without requiring source-code knowledge. 
    \item Prior root cause analysis assumes \vrf has the entire execution context, including an initial memory snapshot. 
    As discussed earlier, this assumption does not hold for \CFA.
    To address this, \acron takes a different approach:
    upon detecting corrupted control data (e.g., return addresses or indirect call targets), \acron performs an additional backward definitions analysis to reconstruct a \emph{symbolic} memory state leading to the corruption.
    Since the recovered state is not fully concrete, existing root cause analysis techniques cannot be directly applied. \acron overcomes this by extending prior work’s forward symbolic data flow analysis to handle symbolic information.
    %% adam- changed "causing" to "leading to" since the symb. memory state we identify is the last valid state 
    %Thus, \acron presents a novel approach to enable root cause analysis without requiring an initial memory snapshot.
    % OAK: technically, do we even need an initial (concrete) memory snapshot at all for root cause analysis? Or do we recover the snapshot (concrete or symbolic) during the backward pass? I think the key missing here is existing work uses this snapshot but we dont need it. Is it because we can reconstruct it from CFA or simply we don't need it or we just need a symbolic version of the snapshot.
    %% adam: i see what you mean. For us, it is because we just need a symbolic version of the snapshot (the start of CFSlice)
    % OAK: \cflog hasnt been introduced yet...
    %Instead, \acron uses backward definitions analysis to determine a trace midpoint at which forward symbolic data-flow analysis begins. This combined approach allows \acron to identify the root cause using only \cflog and the attested binary.
    %
    Based on this analysis and the detected vulnerability, \acron automatically generates binary patches and validates their effectiveness.
    %Through the processes available in \acron, \vrf can do the following: detect a malicious program path present in the run-time evidence, identify the corrupted control data responsible for deviating from a valid path, symbolically execute a program slice to identify the exact memory instruction utilized during the buffer overflow, automatically generate a patch for the compromised memory instruction, and validate the patch effectiveness against attempts to corrupt the same control data.
    \item We develop an open-source prototype of \acron, supporting TI MSP430 and ARM Cortex-M MCU binaries~\cite{repo} and run-time evidence generated by existing open-source run-time attestation RoTs (based on Trusted Execution Environments (TEEs)~\cite{traces} and custom hardware~\cite{acfa}).
\end{myenumerate}

\section{Background}\label{sec:bg}

\subsection{Remote Attestation}\label{subsec:ra}
\RA is a challenge-response protocol between two parties: a remote \vrf and a potentially compromised \prv.
It allows \vrf to remotely assess \prv trustworthiness by measuring the content of \prv program memory.
As depicted in Fig.~\ref{fig:ra}, a typical \RA interaction involves the following steps:

\begin{myenumerate}
    \item \vrf requests attestation from \prv by sending a cryptographic challenge $Chal$.
    \item Upon receiving $Chal$, \prv computes a measurement over its program memory and $Chal$ to produce report $H$.
    \item \prv sends the report $H$ back to \vrf.
    \item Upon receiving $H$, \vrf checks $H$ against the expected value to determine if \prv has been compromised.
\end{myenumerate}

Early ``software-based'' attestation~\cite{kennell2003establish,swatt,pioneer,sake} methods performed step (2) without relying on secrets and instead incorporated measurable architectural side-effects (e.g., execution/response time) along with the measurement of program memory. To make attestation more suitable to remote wireless connections with unpredictable network latency, subsequent methods required \prv to respond with an authenticated integrity token --- e.g., a message authentication code (MAC) or a digital signature --  computed over program memory using a secret key (either a pre-shared symmetric key or the private key for a public key \vrf knows).
Since threat models of \RA (and of \CFA consequently) assume that \prv is susceptible to full software compromise, the secret key used in this operation must be securely stored by \prv's RoT. Secure storage for the \RA secret key implies some level of hardware support for the RoT implementation, ensuring that the key is unmodifiable and inaccessible to any untrusted software running on \prv.
Therefore, ``hardware-based'' approaches incorporate dedicated hardware RoT for storing secret keys and computing the measurement~\cite{copilot,checkmate_att,tpm_attest,sacha,Sancus17}. Since such hardware is too costly for low-end MCUs, recent approaches have proposed ``hybrid'' RoTs~\cite{vrased,smart,tytan}, which use hardware for storing secret keys and trusted software for performing the authenticated measurement.

\subsection{Code Reuse and Control Flow Attacks}\label{subsec:attacks}
%%% condensed
Control flow attacks are made possible due to an exposed memory vulnerability, such as common and pervasive instances like buffer overflow and use-after-free vulnerabilities~\cite{mitre-cwe-top-25-2024}.
Buffer overflows are common in embedded systems software since it is often written in unsafe memory languages like C/C++. For example, recent CVE's~\cite{cve_2021_35395,cve_2020_10019,cve_2020_10023} have disclosed buffer overflows exploited for arbitrary code execution in IoT devices~\cite{realtek_vuln_report} and embedded real-time operating systems~\cite{zephyr_project}.
Use-after-free is another class of vulnerabilities that can lead to full software compromise.
As it is not uncommon for embedded software and RTOSes to use dynamic (i.e., heap) allocation~\cite{masmano2003dynamic,ramakrishna2008smart,deligiannis2016adaptive}, they can also be vulnerable to use-after-free attacks, which is evidenced by recently disclosed CVE's~\cite{cve_2017_14201,cve_2021_0920}.
Since these attacks do not require code modifications, they would remain undetected by classic \RA mechanisms.

\begin{figure}[t]
    \begin{center}
        % \resizebox{\columnwidth}{!}{
        \begin{tikzpicture}
            \node[text width=3cm] at (3.65,1.75) {Verifier (\vrf)};
            \fill[blue!40!white] (3,0) rectangle (3.25,1.5);
            \node[text width=3cm] at (2.1, 0.1) {(4) Verify Report};
            
            \draw[->, thick] (3.3, 1.0) -- node[above] {(1) Request} (5.45,1.0);
            \draw[->, thick] (5.45, 0.2) -- node[above] {3) Report} (3.3, 0.2);
            
            \node[text width=3cm] at (6.5,1.75) {Prover (\prv)};
            \fill[blue!40!white] (5.5,0) rectangle (5.75,1.5);
            \node[text width=3cm] at (7.35,0.6) {(2) Measure state};
        \end{tikzpicture}
    \end{center}
    \vspace{-1.5em}
    \caption{\RA interaction}
    \label{fig:ra}
    \vspace{-1.5em}
\end{figure}

%%%% old
% Another memory vulnerability that often leads to control flow attacks is use-after-free~\cite{cwe-416} (a concrete example leading to control flow hijacking is presented in Appendix~\ref{apdx:use-after-free}).Use-after-frees were ranked as one of the most dangerous software weaknesses by MITRE~\cite{mitre-cwe-top-25-2024}. Embedded libraries/firmware that use dynamic allocation~\cite{masmano2003dynamic,ramakrishna2008smart,deligiannis2016adaptive} can be equally vulnerable use-after-free vulnerabilities~\cite{cve_2017_14201,cve_2021_0920}. An example of use-after-free leading to arbitrary code execution in embedded platforms includes CVE-2017-14201~\cite{cve_2017_14201} that affected Zephyr RTOS~\cite{zephyr_project}. Whether launched by a buffer overflow or a use-after-free, such attacks do not require code modifications. Thus, they would remain undetected by classic \RA mechanisms.

\subsection{Run-time Attestation and Auditing}\label{sec:cfa}

Run-time Attestation and Auditing are a set of security services that aim to generate and guarantee the delivery of accurate and authentic evidence of a \prv's run-time behavior to a remote \vrf, thus reflecting to \vrf whether any control flow attack has occurred. 
One type of run-time attestation is Control Flow Attestation (\CFA)~\cite{tinycfa,oat,litehax,atrium,scarr,recfa,cflat,geden2019hardware,iscflat,blast,enola}, which extends \RA to detect run-time attacks in addition to illegal binary modifications. Upon receiving a request to execute some software function along with a \RA challenge from \vrf, \prv executes the requested function, and a RoT in \prv records a trace of this execution into protected memory.
In the most fine-grained case, the trace is a control flow log (\cflog) of all control flow transfers that have occurred during the requested execution.
After producing \cflog, \prv performs the authenticated \RA integrity measurement over the \cflog, the received challenge, and its own program memory to produce a timely and authenticated report. This report attests to the static state of \prv's program binary and the run-time state within this timely execution. When applicable, data inputs -- in Data Flow Attestation (\DFA)~\cite{oat,dialed,litehax} -- or produced outputs~\cite{apex,asap} can also be included in this proof to bind obtained results to the proper execution of respective software functions on expected inputs.

Existing \CFA techniques use either (1) binary instrumentation along with TEE support; or (2) custom hardware modifications to generate \cflog by detecting and saving each branch destination to a dedicated and protected memory region. 
For techniques that use binary instrumentation, a pre-processing phase modifies the binary so that all branch instructions (e.g., \texttt{jumps}, \texttt{returns}, \texttt{calls}, etc.) are prepended with additional calls to a TEE-protected trusted code. Once called, the trusted code appends \cflog to update the trace with the current branch destination. In hardware-based techniques, custom hardware interfaces with the MCU core to detect branches 
% as they are executed by their instruction opcode. Once detected, the custom hardware monitors the program counter to determine the branch destination. This destination is
and records their destinations to a reserved memory region.
% by the custom hardware in parallel to the software's execution. 
%Finally, \prv transmits \cflog to \vrf along with the produced authentication token (i.e., the MAC/signature result). In possession of the attested binary and \cflog, \vrf can determine if \prv execution occurred as intended and detect control flow attacks (in addition to binary modifications).
%
%Since the RoT was typically used for building \cflog and generating the attestation evidence, \CFA was limited by the fact that a fully compromised \prv could ignore the protocol, refusing to fully produce/deliver an attestation evidence that indicates a compromised state.
Recent work~\cite{acfa,traces} has also created mechanisms to implement reliable communication with \vrf as a part of the \CFA RoT, enabling \textit{run-time auditing} by ensuring evidence delivery even in the presence of malware that infects \prv and refuses to deliver evidence about its actions/presence.

\section{\CFA Evidence and its Utility}\label{sec:classify}

\subsection{Evidence Types}\label{subsec:types}

The various realizations of \CFA RoTs (see Sec.~\ref{sec:bg}) create and manage \cflog differently, presenting trade-offs between the storage/communication cost and ease of verification/analysis by \vrf. We classify existing methods in three types: hash-based evidence (\textbf{E1}), verbatim evidence (\textbf{E2}), and hybrid evidence (\textbf{E3}). % Here, we also discuss their relative storage/communication cost and introduce their feasibility of verification/evidence analysis (discussed further in Section~\ref{subsec:detecting}).
\edit{Compared to Ammar et al.~\cite{sok_cfa_cfi}, we further offer a more detailed analysis of each type’s suitability for advanced analysis.}
% including their ability to pinpoint and remediate root causes concretely.}

\begin{figure}[t]
    \centering
    \includegraphics[width=\columnwidth]{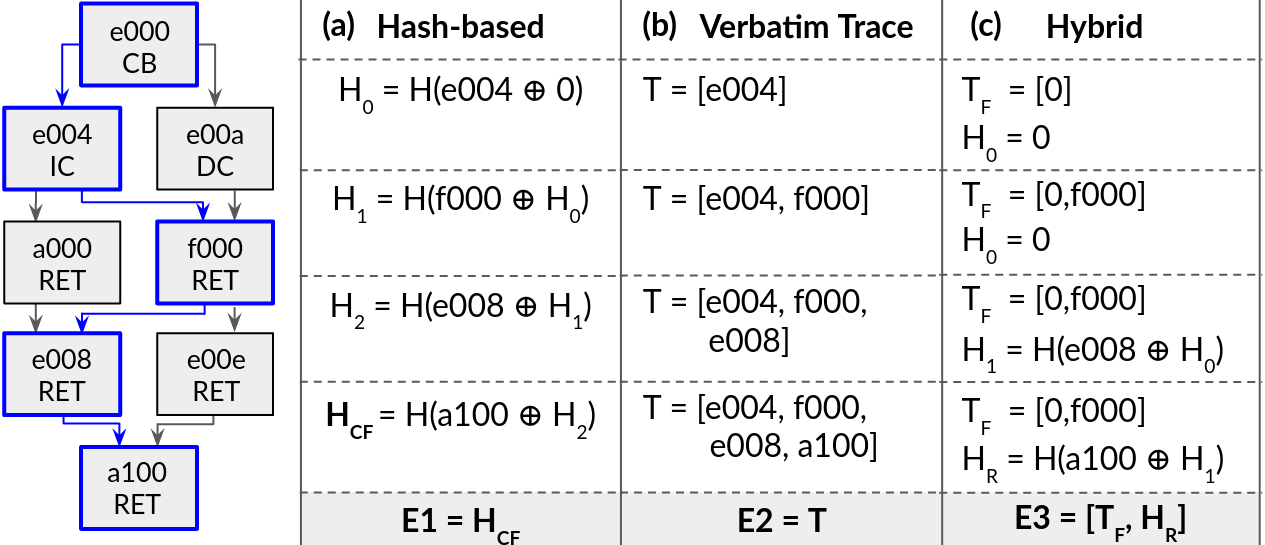}
    \vspace{-2em}
    \caption{A conditional branch (CB), indirect/direct calls (IC/DC), and return (RET) in (a) hash-based, (b) verbatim, and (c) hybrid evidence.}
    \label{fig:gen}
    \vspace{-1em}
\end{figure}

% \subsubsection{Hash-based Evidence (E1)}
\textbf{Hash-based Evidence (E1).}
As control flow transfers occur, branch destination addresses are accumulated into a single hash chain. This is performed by first accumulating the current destination address with the previous hash and then hashing the accumulated result~\cite{cflat,lofat,atrium}, as exemplified in Fig.~\ref{fig:gen}.a. After the first branch occurs, its destination \texttt{0xe004} is accumulated into the hash. Since this is the first branch destination, it is accumulated with an initial constant to obtain $H_0$. Upon the second branch instruction, its destination address \texttt{0xf000} is accumulated with $H_0$ to obtain $H_1$. This process continues until the end of the requested execution generating a unique value that represents the path. Although simple control flows produce a single hash in \textbf{E1}, programs with more complex control flows (including those with nested loops and branching) might require a set of hash digests~\cite{cflat,atrium}. 
This approach yields minimal storage/transmission overheads (i.e., the fixed size of a hash digest).
However, \vrf must perform its analysis solely based on the final hash digest and the executable binary.
%%However, \vrf's ability to analyze the evidence is limited. Considering that the control flow path is not included in the evidence, the analysis would require \vrf to determine the path that leads to the received hash digest. The complexity of this task grows exponentially with path sizes, leading to the intractable path explosion problem\todo{~\cite{}}.

% \subsubsection{Verbatim Evidence (E2)}
\textbf{Verbatim Evidence (E2).}
On the other side of the spectrum, more recent \CFA techniques advocate for storing and transmitting \cflog in its entirety \cite{tinycfa,litehax,scarr,acfa,recfa,traces}. As depicted in Fig.~\ref{fig:gen}.b, the full sequence of destination addresses is maintained in a buffer. Therefore, this approach requires further (lossless) compression or is limited to small and self-contained operations to avoid filling the dedicated storage too quickly. A common strategy is to log simple loops (those without internal branches) with their address once and a count/condition pertaining to the number of iterations performed~\cite{acfa,tinycfa,recfa,traces}. Another reduction is obtained by recording statically defined branch destinations as a single bit~\cite{litehax}. Since static destinations can be determined from the binary alone, branch instructions with statically defined destinations can be recorded as a single bit (1/0) to determine if the branch was taken or not. Branches with dynamically defined destinations (e.g., returns and indirect calls) are still recorded with their full addresses. If \cflog remains larger than available storage despite these optimizations, it can be split into a set of multiple fixed-size slices and streamed to \vrf, as suggested in~\cite{scarr,acfa,traces}. Despite the aforementioned challenges, {\bf E2} provides \vrf with full control flow path information for analysis.
    
% \subsubsection{Hybrid Evidence (E3)}
\textbf{Hybrid Evidence (E3).}
Some techniques implement a combination of {\bf E1} and {\bf E2}. OAT~\cite{oat} and ARI~\cite{ari} record forward edges {\it verbatim} and accumulate backward edges (returns) into a single hash chain. As shown in Fig.~\ref{fig:gen}.c, the RoT in \prv maintains a trace of forward edges ($T_F$) and a hash chain of returns ($H_R$). Since \texttt{0xf000} is the destination of an indirect call, its full address must be recorded whereas the first destination can be represented by a single bit `0' encoding the `not-taken' conditional branch address. After the attested execution ends, $T_F$ and $H_R$ are sent to \vrf.

\subsection{Detection \& Remediation via E1, E2 \& E3}
\label{sec:detecting}

\begin{figure*}[t!]
    \centering
    \includegraphics[width=0.85\textwidth]{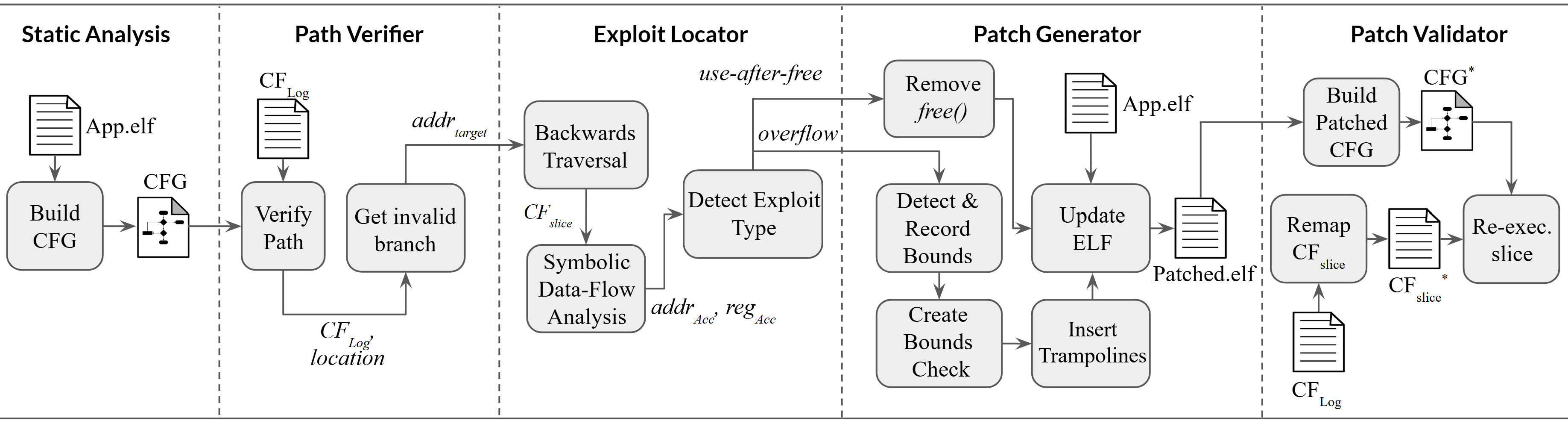}
    \vspace{-1.25em}
    \caption{\acron sub-modules and internal workflow}
    \label{fig:workflow}
    \vspace{-1em}
\end{figure*}

Consider an attack detection strategy in which \vrf constructs the CFG of the executable binary, i.e., as suggested in \cite{cflat,acfa,lofat,atrium,scarr,recfa}. An initial set of \textit{invalid paths} is defined as forward edges that do not exist in the statically constructed CFG. As valid indirect targets (e.g., return addresses or indirect calls) are over-approximated using the static CFG~\cite{burow2019sok}, \vrf may also use \cflog along with the program's binary to emulate the execution of the program locally. In this process, a Shadow Stack (SS) can be emulated locally to verify backward edges (return destinations) and data flow (DF) analysis can be employed to determine a set of valid call sites for each indirect call, updating the CFG accordingly.
 We observe feasibility of this verification strategy varies greatly depending on evidence types {\bf E1}, {\bf E2}, and {\bf E3}.
 
For \textbf{E1}, with the only evidence available being a hash digest, \vrf must infer which path has resulted in the digest. To that end, it must generate each possible legal path based on the CFG and use an emulated SS to accumulate the path's hash during each step of the CFG traversal. However, the complexity of this task grows exponentially with the number of branches, leading to the intractable path explosion problem~\cite{aliasing,baldoni2018survey}. Considering anomalous/illegal paths, the analysis is even more challenging. \vrf would need to map the hash digest to the exact illegal path. Illegal paths do not exist in the CFG and thus must then be enumerated directly, making it effectively impossible to learn about the exploit behavior from this type of evidence (this precludes any analysis of exploit root causes, as discussed below).

Evidence types \textbf{E2} and \textbf{E3} allow \vrf to use reported traces to traverse the CFG, checking the reported destination against valid destinations. When there is a mismatch between the reported destination and the set of valid destinations for that branching instruction, \vrf determines that \prv executed an invalid path. With both \textbf{E2} and \textbf{E3}, \vrf is able to emulate an SS of return addresses. For verifying the integrity of returns in \textbf{E2}, \vrf checks each return destination individually as they are reached in the path traversal. The expected return address is popped from the SS and compared to the return address reported in \cflog, and an invalid path is detected when there is a mismatch. For \textbf{E3}, however, \vrf must recompute all returns reached into the final hash chain digest and compare it to the one reported in \cflog. If there is a mismatch of hashes, an invalid path is detected due to one of the return addresses.
We note that verification based on {\bf E3} is only able to determine that some return address was corrupted.
Since actual return addresses are not logged {\it verbatim}, the path explosion problem still applies when attempting to determine which return in the chain was corrupted, preventing accurate root cause analysis by \vrf.

%%Ivan: This section seems unecessary
%\section{Overview}
%Prior works focus on a secure implementation of the \CFA RoT on \prv in order to ensure that the generated evidence is correct and reliably delivered to \vrf. However, the task of analyzing valid and invalid paths from the reported evidence has largely been overlooked in prior work. This is possibly due to the fact that there are sometimes infinite possible valid paths or that it is challenging to learn what happens when only the \cflog and program binary are available. This challenge is apparent even for simple embedded systems with less complex programs. 

%In Section~\ref{sec:effective}, we study the extent to which information can be learned from the various forms of control flow evidence, concluding that evidence that might be more optimal from \prv point of view is less optimal for \vrf to go beyond compromise detection. Given these observations, we present \acron in Section~\ref{sec:design} to support \vrf further analysis \cflog to determine if it represents a valid path. Next, it goes beyond detection to identify the exploited control instruction, pinpoint the exploited memory instruction, and provide a patched version of the identified code segments to be deployed through a software update. \acron achieves this when the source code is not available (i.e., over the binary executable alone).

The discussion above highlights that \vrf's ability to learn an attack's behavior and its root cause varies. Put simply, \textbf{E1} only allows \vrf to learn if the reported path is invalid. However, it is impossible to learn what the invalid control flow path was or how it was exploited.

With {\bf E2}, an invalid path is determined when a forward-edge target does not match possible successors in the CFG or when a return address in \textbf{E2} does not match the value popped from the SS. In either case, \vrf can determine the exact indirect call or return exploited since all of these destinations are included in \cflog. 

In {\bf E3}, since \vrf only receives a hash chain of all return addresses, it cannot determine which return was corrupted or its destination address after the corruption. % When a return is reached during validation of \textbf{E3}, \vrf pops the expected return address from the SS, updates the expected hash $H_R'$, and then travels to the popped address to continue verifying the trace.
As a consequence, \vrf cannot make sense of subsequent forward edges in \cflog because it cannot verify that \prv has followed the expected control flow transfers (as a previously exploited return could have gone anywhere). At this point, continuing to traverse the CFG yields no additional information. Thus, \vrf can not learn anything more about the execution, resulting in limited analysis capability beyond detecting that some control deviation occurred. In sum, similar to {\bf E1}, {\bf E3} results in \vrf being unable to examine exploit root causes. 

It follows from the discussion above that \textbf{E2} is the most adequate to support auditing and exploit root cause analysis by \vrf based on \cflog. Based on this conclusion, Sec.~\ref{sec:design} presents \acron to showcase how \vrf analysis can be performed based on a \cflog when it includes a {\it verbatim} sequence of control flow transfers.

% \begin{figure*}[t!]
%     \centering
%     \includegraphics[width=0.85\textwidth]{figs/workflow.png}
%     \vspace{-1.25em}
%     \caption{\acron sub-modules and internal workflow}
%     \label{fig:workflow}
%     \vspace{-1em}
% \end{figure*}

% \vspace{-3em}
\section{\acron}
\label{sec:design}

\acron showcases \cflog-based root cause analysis and remediation with {\it verbatim} control flow traces (\textbf{E2}) and without the source-code of the attested program.
Fig.~\ref{fig:workflow} shows \acron workflow consisting of five internal modules: Static Analysis, Path Verifier, Exploit Locator, Patch Generator, and Patch Validator.
\acron takes as input an application binary (\appbin) and a verbatim \cflog (\textbf{E2} evidence type). The notation used in this work is summarized in Table~\ref{tab:notation}.

%In an offline phase, \vrf builds the CFG of the application binary (\appbin). In the case of TEE-based \CFA techniques, this offline phase would also instrument \appbin creating a modified version for logging control flow transfers. During the online phase, \vrf receives the \cflog as a part of \textbf{E2} from \prv. These three objects are used by \acron as inputs to perform the remaining analysis.

%%% MINIPAGE VERSION 
% \begin{figure*}[!th]
% \begin{minipage}{0.30\textwidth}
%     \centering
%     \includegraphics[width=\columnwidth]{figs/cfg-cflog-example.png}
%     \caption{Example CFG and \cflog}
%     \label{fig:cfg}
% \end{minipage}
% \hfill
% \begin{minipage}{0.33\textwidth}
%     \centering
%     \includegraphics[width=\columnwidth]{figs/back-trace-ret.png}
%     %\vspace{-0.5em}
%     \caption{Backward Traversal for a Return Address}
%     \label{fig:back-trace-ret}
%     %\vspace{-0.5em}
% \end{minipage}
% \hfill
% \begin{minipage}{0.34\textwidth}
%     \centering
%     \includegraphics[width=\columnwidth]{figs/back-trace-ic.png}
%     %\vspace{-0.5em}
%     \caption{Backward Traversal for an Indirect Call}
%     \label{fig:back-trace-ic}
%     %\vspace{-0.5em}
% \end{minipage}
% \end{figure*}

\subsection{Static Analysis and Path Verifier}
\label{subsec:static-analysis}

\begin{figure*}[t]
    \centering
    \begin{minipage}{0.27\textwidth}
        \centering
        \includegraphics[width=\linewidth]{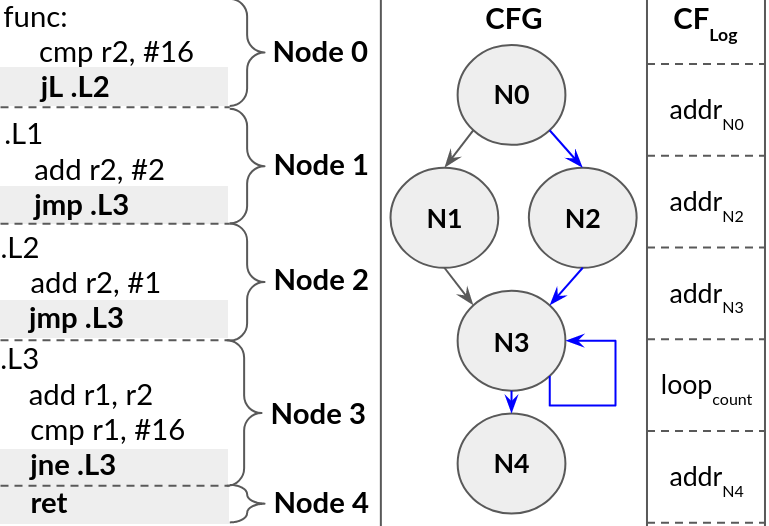}
        \vspace{-2em}
        \caption{CFG and \cflog}
        \label{fig:cfg}
    \end{minipage}
    \hfill
    \begin{minipage}{0.35\textwidth}
        \centering
        \includegraphics[width=\linewidth]{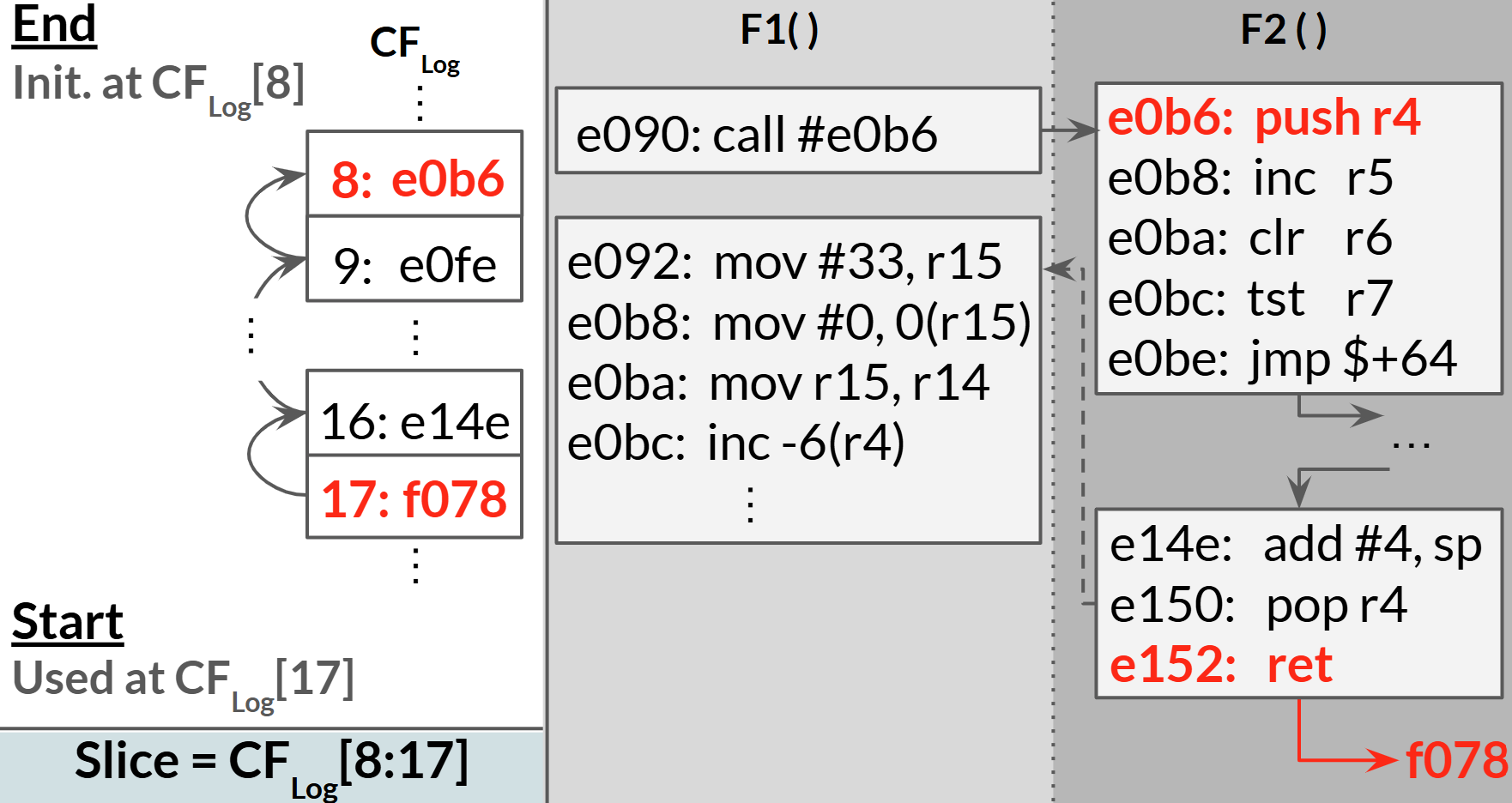}
        \vspace{-2em}
        \caption{Backward Traversal for a \texttt{ret}}
        \label{fig:back-trace-ret}
    \end{minipage}
    \hfill
    \begin{minipage}{0.37\textwidth}
        \centering
        \includegraphics[width=\linewidth]{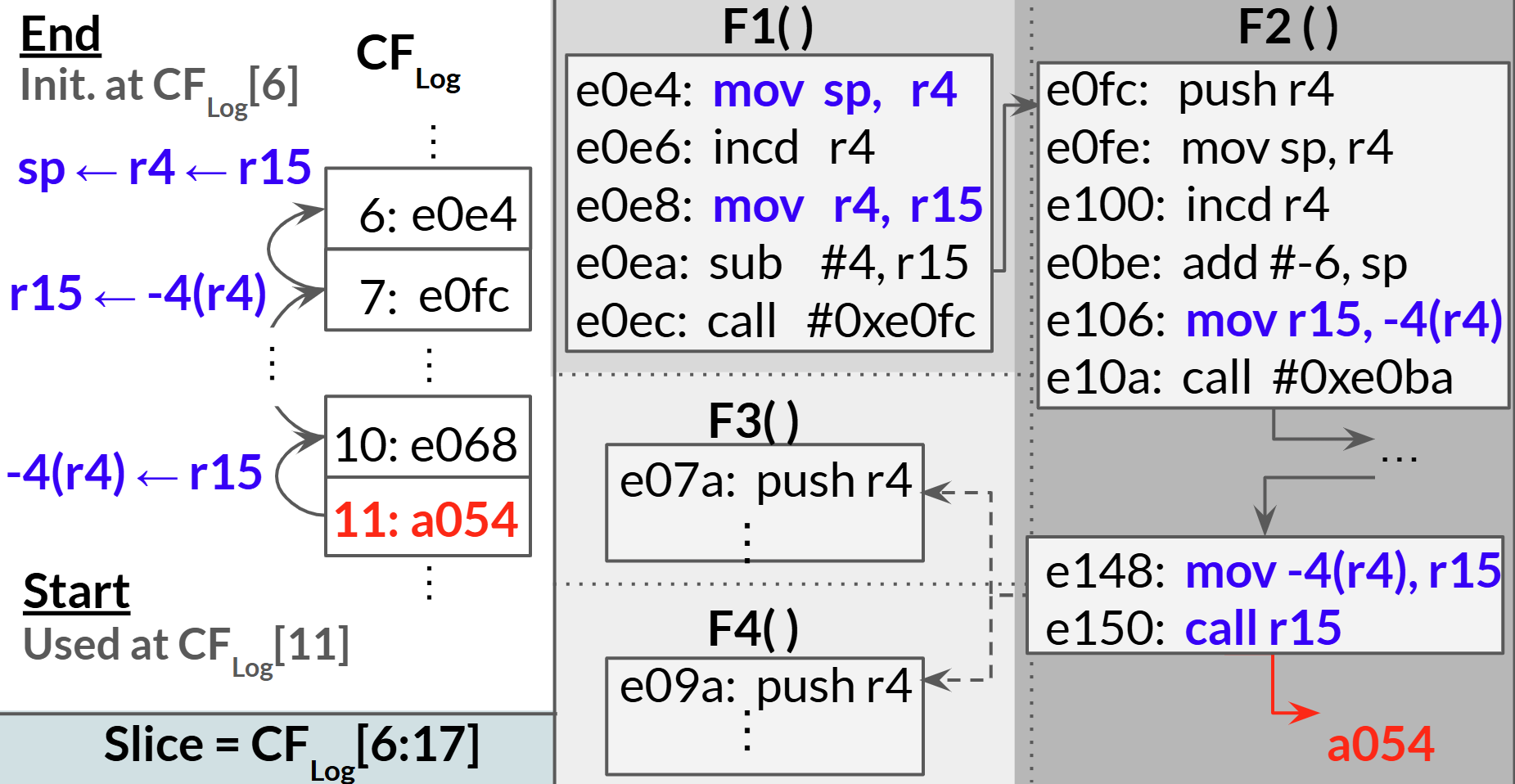}
        \vspace{-2em}
        \caption{Backward Traversal for an \texttt{call}}
        \label{fig:back-trace-ic}
    \end{minipage}
    \vspace{-1em}
\end{figure*}

% OAK: what's the point of saying this? They may respond: why putting these stages here then?
The first two stages of \acron are well studied from prior works~\cite{sok_cfa_cfi}. 
The first module of \acron performs static analysis of the application binary (\appbin) to produce the CFG used by subsequent modules. 
Each CFG node represents a set of contiguous (non-branching) instructions starting with the previous branch destination and ending with the next branch, as shown in Fig.~\ref{fig:cfg}.
This module is executed once per \appbin. The remaining modules are executed for each received \cflog. 
% Since \acron also can verify path validity from any \CFA evidence, this offline phase optionally can include computing a set of valid hashes ($H_{set}$) that encompasses all valid paths (if possible to reach) within the CFG for \textbf{E1} verification.

% \subsection{Path Verifier}
% \label{subsec:locating_ctrl_instr}

\acron verifies the contents of the verbatim \cflog, the contents of which are assumed to have the format depicted in Fig.~\ref{fig:cfg}. 
\acron interprets each \cflog entry as either: (1) a branch destination (denoted $addr_{Ni}$ in Fig.~\ref{fig:cfg}) or (2) a loop counter (denoted $loop_{count}$ in Fig.~\ref{fig:cfg}), indicating the number of times that a CFG node in the previous entry has executed. 
After receiving \cflog, Path Verifier uses this information to determine whether \cflog constitutes a valid control flow path by traversing the CFG and SS, as described in Sec.~\ref{sec:detecting}.
%It may also contain a loop counter,  optimizations for static loops~\cite{tinycfa,acfa}, in which a loop counter ($loop_{count}$ in Fig.~\ref{fig:cfg}) is logged with an identifying symbol instead of repeated instances of the same destination address.
Upon detecting an invalid transfer, it proceeds to analyze 
the type of control flow instruction (i.e., \texttt{ret} or indirect \texttt{call}) whose destination address is being corrupted.
This is done by first locating the node in the CFG that aligns with the last valid transfer in \cflog.
The corrupted instruction is assigned as the last instruction of this CFG node.
Using $App.elf$, the corrupted instruction and target address (\ctrldata) are identified as either a return address or an indirect branch target.
This module outputs the corrupted instruction and \ctrldata to the next module of \acron.

\begin{table}[t!]
    \centering
    \caption{Notation Summary}
    \label{tab:notation}
    \vspace{-1.5em}
    \resizebox{\columnwidth}{!}{
    \small
    \begin{tabular}{|c|p{\linewidth}|}
         \hline
         \textbf{Symbol} & \textbf{Description} \\
         \hline
         \cflog & Verbatim CFA evidence (type \textbf{E2})\\
         \hline
         \textit{App.elf} & The binary of the attested application \\
         \hline
         \textit{CFG} & The Control Flow Graph of \textit{App.elf} \\
         \hline
         \textit{SS} & Shadow Stack of return addresses \\
         \hline
         $loop_{count}$ & a special entry in \cflog to optimize static loops (i.e., loops with no internal branches)~\cite{tinycfa,acfa}; denotes total times the previous \cflog entry repeated \\
         \hline
         \ctrldata & The control data (e.g., return address, indirect jump/call target) that was corrupted\\
         \hline
         \slice & A \cflog slice from \ctrldata's initialization point its corrupted usage. \\
         \hline
         $base$ & The symbol (\textit{i.e., the base} address or register) storing \ctrldata \\
         \hline
         $MemMap$ & address-value mapping for in-use memory addresses during Symbolic DF Analysis. \\
         \hline
         $RegMap$ & register-value mapping for in-use registers during Symbolic DF Analysis. \\
         \hline
         $FreeList$ & list of all freed pointers during Symbolic DF Analysis \\
         \hline
         $X$ & special symbolic value assigned to $base$ during Symbolic DF Analysis. \\
         \hline
         $addr_{Acc}$ & the address of the memory instruction used to corrupt \ctrldata \\
         \hline
         $reg_{Acc}$ & the register $addr_{Acc}$ dereferences to corrupt \ctrldata\\
         \hline
         \multirow{2}{*}{$reg_{Acc}^{Init}$} & the initial value of $reg_{Acc}$ before it is modified to dereference/corrupt \ctrldata. In the context of buffer overflows, it holds the start address of the overflown buffer. \\
         \hline
         $addr_{lower}$ & the instruction address that sets the lower bound of the buffer\\
         \hline
         $addr_{upper}$ & the instruction address that sets the upper bound of the buffer\\
         \hline
    \end{tabular}
    }
    \vspace{-1em}
\end{table}

\subsection{Exploit Locator}
\label{subsec:locating_mem_instr}

% This module pinpoints exploit sources in three phases.

% \subsubsection{Phase 1. Backwards Traversal}\label{sec:backwards_traversal}
\textbf{Phase 1. Backwards Traversal.}\label{sec:backwards_traversal}
Given the corrupted instruction from the Path Verifier, this phase attempts to find:
\begin{myenumerate}
    \item \slice: the section of \cflog relevant to this particular exploit, including the \slice entries between the time when memory storing \ctrldata was assigned with the last pre-corruption value and the time that the corrupted branch executes causing the detected violation.
    \item $base$: the memory address storing \ctrldata at the beginning of \slice. 
\end{myenumerate}
The procedure for determining \slice and $base$ varies based on the type of \ctrldata. This is shown in Fig.~\ref{fig:back-trace-ret} and Fig.~\ref{fig:back-trace-ic}.% and described below:

% \begin{figure}[t]
%     \centering
%     \includegraphics[width=0.8\columnwidth]{figs/cfg-cflog-example.png}
%     \caption{Example CFG and \cflog}
%     \label{fig:cfg}
% \end{figure}

% \begin{figure}[t]
%     \centering
%     \includegraphics[width=0.99\columnwidth]{figs/back-trace-ret.png}
%     \vspace{-0.5em}
%     \caption{Backward Traversal for a Return Address}
%     \label{fig:back-trace-ret}
% \end{figure}

% \begin{figure}[t]
%     \centering
%     \includegraphics[width=0.99\columnwidth]{figs/back-trace-ic.png}
%     \vspace{-0.5em}
%     \caption{Backward Traversal for an Indirect Call}
%     \label{fig:back-trace-ic}
% \end{figure}

% \begin{myitemize}
    % \item 
    {\bf Corrupted returns.} Return addresses are initialized when pushed onto the stack during function calls, and they remain unmodified until the respective returns. 
    Consequently, any corruption to a return address must have occurred between the function call and the return instruction.
    Therefore as a first step, \appbin is inspected to determine the start address of the function containing the corrupted return. 
    In the example from Fig.~\ref{fig:back-trace-ret}, Path Verifier detects that the 17th \cflog entry is invalid, corresponding to the corrupted return instruction at address \texttt{0xe152} and \ctrldata pointing to \texttt{0xf078}.
    This module then inspects the disassembled instructions of \appbin and identifies that function \texttt{F2} contains this corrupted instruction.
    Then, it saves the start address of \texttt{F2}, \texttt{0xe0b6}, as the target of the backward traversal. 
    A backward traversal of \cflog is performed to find the last call to \texttt{F2}, which happens to be at the 8th \cflog entry.
    As a result, this phase returns \slice consisting of entries 8 to 17, i.e., \texttt{\cflognospace[8:17]}.
    In addition, since return addresses are always initialized on the stack, $base$ can be found in the stack pointer register (\texttt{sp}) at the beginning of \slice. Hence, this phase also outputs $base$ to be \texttt{sp}.

    % \item
    {\bf Corrupted forward edges.} 
    Indirect forward edge targets (indirect jumps/calls) are stored in a register operand at the time of their use.
    To identify \ctrldata's initialization point in this case, \cflog is used for a backward traversal of CFG while repeatedly following its assignments until reaching its initialization.
    For example, in Fig.~\ref{fig:back-trace-ic}, Path Verifier determines that, at instruction address \texttt{0xe150}, \ctrldata is stored in \texttt{r15}.  Thus, this phase traverses instructions backward from address \texttt{0xe150} until it finds the latest assignment of \texttt{r15} at address \texttt{0xe148}. As \texttt{r15} is assigned by \texttt{-4(r4)} at this address, Exploit Locator switches to tracking this new memory instead. 
    This process continues until it encounters the instruction that initializes \ctrldata, which then sets the beginning of \slice to the first \cflog entry that includes this instruction.
    This phase finally assigns $base$ according to the type of the source operand in this instruction:
    \begin{myenumerate}
        \item the stack pointer (i.e., where \ctrldata is declared as a local variable), which is the case in  Fig.~\ref{fig:back-trace-ic}.
        In this example, this phase stops once it finds a match of the source operand with \texttt{sp} at \texttt{0xe0e4}.
        Since this instruction reflects the 6th entry of \cflog and the corrupted instruction happens in the 11th, it returns $\slice = \texttt{\cflognospace[6:11]}$.
        It outputs $base = \texttt{sp}$.
        \item a constant indicating a fixed memory address (i.e., when \ctrldata is declared as a global variable).
        $base$ is set to this fixed address.
        \item a return value from \textit{malloc} (i.e, in the case where \ctrldata is dynamically allocated). Backward traversal stops at the last \textit{malloc} that defined it. Thus, \slice only includes the \textit{malloc} and \textit{free} that potentially lead to the attack. In this case, $base$ is set to the return register from the \textit{malloc}.
    \end{myenumerate}

\textbf{Phase 2. Symbolic DF Analysis.}
\ignore{
At this point, Exploit Locator obtains the following information from \cflog and \appbin:
\begin{myitemize}
    \item \slice: the execution slice representing the start and end of the search area for the memory operation that corrupts \ctrldata
    \item \textit{base}: the symbol (register or fixed memory address) holding the last uncorrupted \ctrldata at the start of \slice.
\end{myitemize}
}
Within \slice determined by Phase 1, Exploit Locator must locate the instruction that corrupts \ctrldata, typically requiring emulation of \slice execution.
Without access to run-time data prior to \slice execution, Exploit Locator emulates this execution using symbolic analysis.
Unlike traditional symbolic execution~\cite{king1976symbolic}, we do not use symbolic analysis to explore different paths.
Instead, we use it to reconstruct all possible data flows during \slice execution~\cite{yagemann2021arcus,yagemann2021automated} and monitor for the change in data flows that result in corruption of \ctrldata.

%while monitoring for the instruction that writes \ctrldata during this execution.  

% Register and memory values are evaluated into either concrete or symbolic values based on the information available in the binary. Any register/memory values that cannot be resolved into a concrete value are assigned a symbolic value.
Algorithm~\ref{alg:symb_exec} details this phase.
The symbolic data flow analysis takes \slice, \textit{base}, \appbin, and CFG as inputs. As a first step, it initializes the execution state using \textit{MemMap}, \textit{RegMap}, and \textit{FreeList}, where:
\begin{myitemize}
    \item \textit{MemMap} represents the set of memory states affected by \slice execution, mapping memory addresses to their values.
    It is initially set to empty, except for addresses storing read-only data that can be determined from \appbin (e.g., data in \texttt{.rodata} segment of \appbin).
    
    \item \textit{RegMap} is a map of registers and their values for registers used in \slice portion of \appbin execution. In the initial state, \textit{Reg} is empty.

    \item \textit{FreeList} is a list of all pointers freed during the \slice portion of \appbin execution. \textit{FreeList} is also empty in its initial state.

    %\item \textit{Instrs} contains a list of instructions extracted from \appbin.
\end{myitemize}

The concrete value of \textit{base} is only available during run-time and cannot be determined from \slice.
Therefore, this phase initializes \textit{base} with a special symbolic value ($X$) and updates the state (\textit{MemMap} or \textit{RegMap}) in which \textit{base} is stored accordingly.
For instance, in Fig.~\ref{fig:back-trace-ret} and~\ref{fig:back-trace-ic}, as \textit{base} is found in \texttt{sp} at the start of \slice, $Reg$[\texttt{sp}] is set to $X$.
%This special symbol $X$ will serve as a means to monitor access of \textit{base} in subsequent steps.
%
%Before starting the symbolic execution, \textit{base} is initialized as a special symbolic value ($X$). This special value will be used to track memory address values that are accessed and resolved to \textit{base} during this process. Finally, the corresponding state data ($Reg$ or $Mem$) structure of \textit{base} is updated according to whether it is a register or memory address. For the examples from Fig.s~\ref{fig:back-trace-ret} and~\ref{fig:back-trace-ic}, the stack pointer would be initialized as $X$ in $Reg$. 

% OAK: after some thoughts, do we really need symbolic execution? Would this still work if we initialize all unknown memory/registers to constant (e.g., 0) or some random numbers?

\begin{algorithm}[t]
\caption{Symbolic Data Flow Analysis.}
\label{alg:symb_exec}
\small
\SetAlgoLined
\SetKwInOut{Input}{Input}
\SetKwInOut{Output}{Output}
\DontPrintSemicolon
\Input{\slice : \cflog slice containing vulnerability; \\ 
       $App.elf$ : The application binary; \\
       CFG: Control flow graph of $App.elf$; \\
       $base$ : symbol holding \ctrldata at \slice start;}

\Output{
        $addr_{Acc}$ : address of illegal memory instruction;\\
        $MemMap$ : memory map containing address-value pairs used in \slice;\\
        $RegMap$ : register map containing register-value pairs used in \slice;\\
        $FreeList$ : list of freed pointers during evaluation of \slice;}
\vspace{2mm}        
$MemMap, RegMap, FreeList\leftarrow \emptyset, \emptyset, \emptyset$\;
$MemMap \leftarrow \mathtt{AddConsts}(App.elf)$\;

\eIf{$base~is~a~Register$}{
    $RegMap \leftarrow RegMap \cup \mathit{\{(base, X)\}}$\;
}{
    $MemMap \leftarrow MemMap \cup \mathit{\{(base, X)\}}$\;
}

\ForEach{$entry \in \slice$}{
    
    \eIf{$entry$ is not $loop_{count}$}{
        $dest_{addr} \coloneqq entry$\;
        $max_{iters} \coloneqq 1$\;
    }{
        $max_{iters} \coloneqq entry$\;
    }
    \ForEach{$i \in \{1, 2, ..., max_{iters}\}$}{
        \ForEach{$instr_{addr} \in CFG[dest_{addr}]$}{
            % $op, src, dest \leftarrow Instr[instr_{addr}]$\;
            $MemMap, RegMap, FreeList \leftarrow \mathtt{eval}(instr_{addr}, MemMap, RegMap, FreeList)\;$ 

            \If{$\exists (a,b) \in MemMap$ s.t. $a = X$}{
                $addr_{Acc} \coloneqq instr_{addr}$\;
                % $reg_{Mal} \coloneqq src$\;
                % $pass \coloneqq False$\; 
                % $\texttt{return}$\xspace$ pass, addr_{Acc}, reg_{Mal}$
                $\texttt{return:}$\xspace$(addr_{Acc},MemMap,RegMap,FreeList)$
            }
        }
    }
}
% $addr_{Acc} \coloneqq \texttt{None}$\;
% % $reg_{Mal} \coloneqq \texttt{None}$\;
% $pass \coloneqq True$\; 
% % $\texttt{return}$\xspace$ pass, addr_{Acc}, reg_{Mal}$\;
% $\texttt{return}$\xspace$ pass, addr_{Acc}$\;
\end{algorithm}

After initialization, this phase
proceeds with symbolic execution of \slice by iterating through each $entry$ in \slice.
Assuming that the current $entry$ corresponds to a branch destination, it locates the CFG node containing this destination and extracts all instructions from it. 
Each instruction is then evaluated based on the current program state:
if the instruction depends on memory or registers not currently present in $MemMap$ or $RegMap$, symbolic values (independent from $X$) are assigned to those memory/registers before executing it; otherwise, the instruction is simply executed symbolically.
As a result of symbolic execution, the evaluation updates $MemMap$ and $RegMap$ to reflect the new execution state. Additionally, \textit{FreeList} is updated whenever an evaluated instruction calls \textit{free()}. When this is detected (by inspecting the instruction type and its destination address), the input parameter for the \textit{free()} is first obtained from the previous instruction. Then, the input parameter and the \texttt{call} instruction's address are added to the \textit{FreeList}.

Next, Exploit Locator uses the evaluation results to determine whether a memory write to $base$ has occurred in the current evaluation.
This is done by checking if the value of $X$ in $MemMap$ has changed.
 % a new value has been inserted to address $X$ in the
If so, it indicates that the current instruction introduces a value (i.e., writes) to memory at $base$, leading to \ctrldata's corruption. Exploit Locator then outputs the address of this instruction as $addr_{Acc}$, along with $MemMap$, $RegMap$, and \textit{FreeList} to the next phase.

When no write to $X$ is detected, it continues with symbolic execution over the remaining \slice entries. We note that this analysis may encounter a \slice entry that is a $loop_{count}$ (as depicted in Fig.~\ref{fig:cfg}). In this case, it repeats symbolic evaluation on the same CFG node for $loop_{count}$ times. 

\textbf{Phase 3. Detecting Exploit Type.}
In its final phase, the Exploit Locator attempts to determine whether the control flow attack resulted from a use-after-free or a buffer overflow, using the output from the previous phase.

% \textbf{Use-after-free.}
The \textit{FreeList} can be inspected to determine whether a use-after-free has occurred. Since the previous phase stopped at the point where \textit{base} was overwritten, the Exploit Locator checks the contents of \textit{FreeList} to see if it contains \textit{base}. If \textit{base} is in \textit{FreeList} and it was illegally modified after it was freed, it is likely that \textit{base} was corrupted due to a use-after-free. In this case, Exploit Locator outputs the candidate attack root cause as a use-after-free.

% \textbf{Buffer Overflows}.
The Exploit Locator assumes that an attack exploits a buffer overflow when \textit{base} is not in \textit{FreeList} and when the CFG node containing $addr_{acc}$ has been executed multiple times in \slice (implying that this CFG node is part of a loop that performs repeated memory writes, eventually overflowing into \ctrldata). In this case, Exploit Locator outputs the attack root cause as a buffer overflow.

\begin{figure*}[t]
    \centering
    \begin{minipage}{0.25\textwidth}
        \centering
        \includegraphics[width=\linewidth]{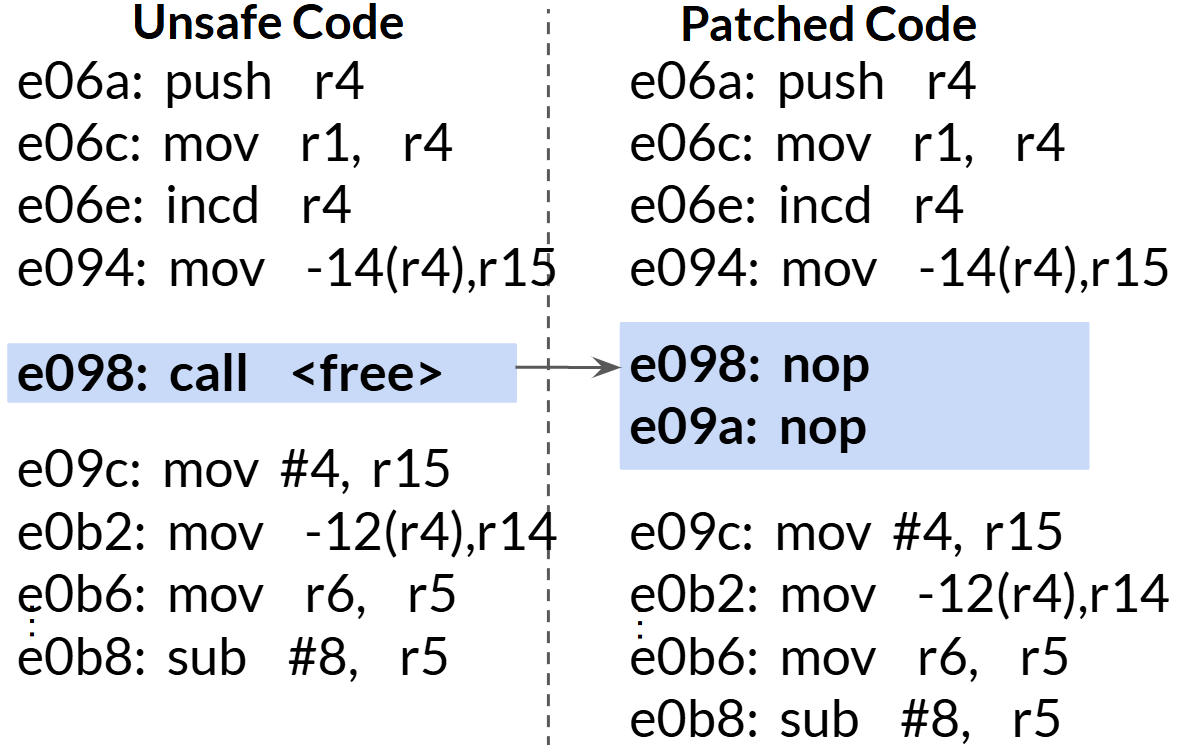}
        \vspace{-0.5em}    
        \caption{Example patch for use-after-free vulnerabilities}
        \label{fig:patch-uaf}
    \end{minipage}
    \hfill
    \vline
    \hfill
    \begin{minipage}{0.73\textwidth}
        \centering
        \includegraphics[width=\linewidth]{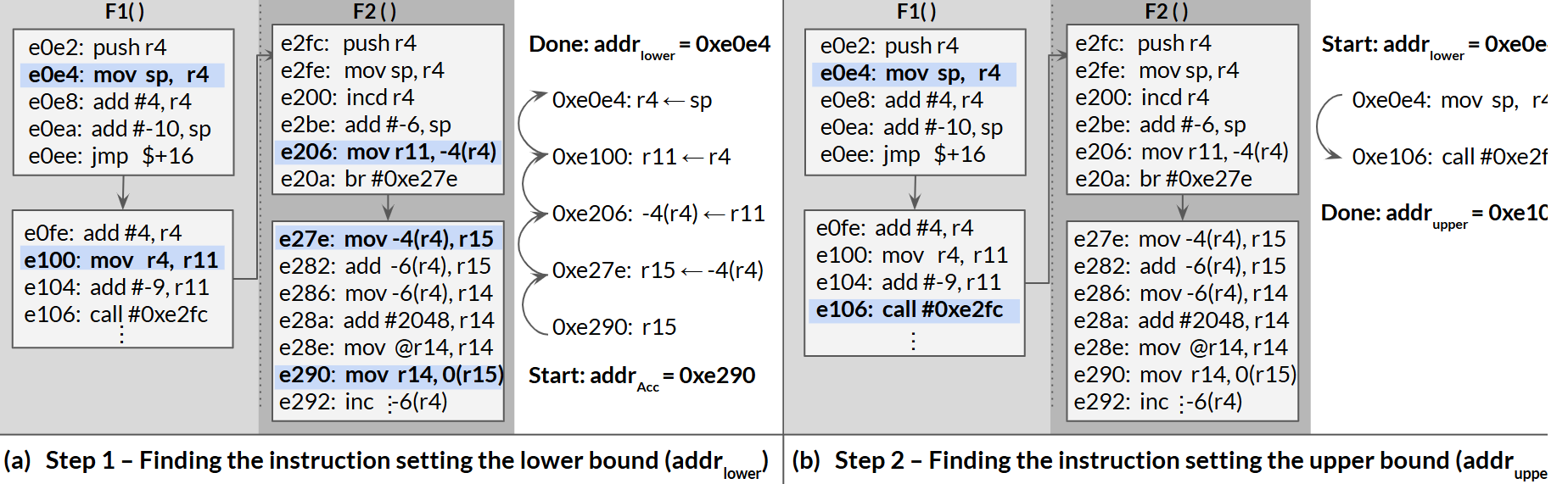}
        \vspace{-2.5em}
        \caption{Backward traversal for determining buffer bounds.}
        \label{fig:determine-buffer-bounds}
    \end{minipage}
    \vspace{-1em}
\end{figure*}

\subsection{Patch Generator}
\label{subsec:patch_generation}

\edit{\acron patching strategy~\cite{osspatcher,vulmet,alice,embroidery} redirects the vulnerable instructions to a secure implementation.}
% The goal of this module is to implement a binary patch that fixes the detected vulnerability. 
The exact steps depend on the detected root cause (either use-after-free or buffer overflow).

% \subsubsection{Patching Use-After-Free}
% \textbf{Patching Use-After-Free.}

% \begin{center}
% \vspace{-0.25em}
% \textbf{Patching Use-After-Free.}
% \vspace{-0.25em}
% \end{center}
\subsubsection{Patching Use-After-Free}

Two root causes lead to a use-after-free: the \textit{free} and the \textit{use} along \slice. 
The \textit{use} of a freed pointer typically occurs when a new object is allocated and when the same freed pointer (i.e., \ctrldata in our case) is indirectly redefined through this new object.
Without source-code semantics (such as variables, structs, pointer types/names), it is challenging to determine whether altering the \textit{use} of \ctrldata will lead to unwanted changes to a program's functionality. Therefore, \acron targets the unintended \textit{free} along slice for patching.

\acron patches the use-after-free by selectively removing the offending \textit{free} from \appbin by replacing it with \texttt{nop} instructions. An example of \acron's use-after-free patch is depicted in Fig.~\ref{fig:patch-uaf}. In this example, call instruction at address \texttt{e098} was identified in the previous step as the call to \textit{free()} in \slice that frees \ctrldata. Call instructions (e.g., \texttt{call} in MSP430 and \texttt{bl} in ARM) take a full address as a parameter and thus span two memory addresses. Therefore, the \texttt{call} is replaced with two \texttt{nop} instructions.

\subsubsection{Patching Buffer Overflows}
% \textbf{Patching Buffer Overflows.}
% \begin{center}
% \vspace{-0.5em}
% \textbf{Patching Buffer Overflows.}
% \vspace{-0.5em}
% \end{center}
% %

Generating a patch for buffer overflow vulnerabilities includes the following tasks:
% This includes two tasks:
\begin{myitemize}
    \item[\textbf{T1:}] Estimate the upper and lower bounds of the buffer that overflows into \ctrldata.

    \item[\textbf{T2:}] Rewrite the instructions in the original code to safely record the bounds.

    \item[\textbf{T3:}] Copy the function containing $addr_{acc}$ to an empty region of memory, wrapping $addr_{acc}$ in new instructions that perform a bounds check.

    \item[\textbf{T4:}]
    Replace the original call to the function from \textbf{T3} with a trampoline to the patched version.
\end{myitemize}

% \begin{center}
% \textbf{Detecting Buffer Bounds}
% \end{center}
\textbf{T1: Estimating Buffer Bounds.}
Buffer overflows are eliminated via bound checks before memory accesses according to the buffer size. From the previous phase, the Exploit Locator identified the offending instruction to be located at $addr_{Acc}$ and that it causes an overflow onto $base$.
This suggests that a buffer exists next to $base$ that must have been overwritten by a loop before the overflow occurs.
It also implies that the offending instruction overflowing to $base$ is the same as that used to write data to the buffer.

With these insights, backtracking to where the destination operand (storing the overflown address) is initialized should provide the start address of the overflown buffer.
Using this address, valid buffer bounds can be estimated to insert bound checks to the binary accordingly. 
This phase aims at finding these bounds before a patch can be generated and applied in the next modules.

% \begin{figure}[t]
%     \centering
%     \includegraphics[width=0.9\columnwidth]{figs/reg-acc.png}
%     %\vspace{-1em}
%     \caption{Backward traversal for determining the access register ($reg_{Acc}$) and its initialization point $addr_{Init}$.}
%     \label{fig:reg-acc}
%     %\vspace{-0.5em}
% \end{figure}

The example of Fig.~\ref{fig:determine-buffer-bounds} depicts the two steps required to obtain the buffer's bounds. The previous phase has detected that the instruction \texttt{mov r14, 0(r15)} triggers a write to \textit{base}, setting $addr_{Acc}$ to \texttt{0xe290}. 
From this instruction, this phase knows that its destination register, \texttt{r15}, stores the write address pointing into the overflown buffer,
and thus labels it as $reg_{Acc}$.
To determine the start address of the overflown buffer, \acron performs a backward traversal from \texttt{0xe290} until it finds the instruction that initializes its lower bound.

As depicted in Fig.~\ref{fig:determine-buffer-bounds}, the buffer's definition is influenced by \texttt{-4(r4)} at \texttt{0xe27e}, which is in turn initialized by \texttt{r11} at \texttt{0xe206}. Tracking of definitions continues until a known pointer value is reached, as described in Sec.~\ref{sec:backwards_traversal}.
In this example, definitions are tracked until \texttt{0xe0e4} since the stack pointer (a known pointer value) has been reached.
Thus, Exploit Locator saves \texttt{0xe0e4} as $addr_{lower}$ (the address defining the lower bound of the exploited buffer) and saves the value of $reg_{Acc}$ (\texttt{sp} in this example) as the lower bound of the overflown buffer at this instruction.
% and \texttt{r11} as $reg_{Acc}^{Init}$ that holds the start address of the overflowed buffer.

After this step, \acron looks forward in \slice to find an instruction that establishes an upper bound for $reg_{Acc}$ used at $addr_{lower}$. For example, since $reg_{Acc}$ is the stack pointer in Fig.~\ref{fig:determine-buffer-bounds}, \acron stops at the next \texttt{call} instruction at address \texttt{0xe106}, as the stack pointer value before this \texttt{call} will determine the upper bound of the current stack frame, providing an approximation of the buffer's upper bound. Therefore in this example, the address \texttt{0xe104} (the instruction before the \texttt{call}) is saved as the location that defines the upper bound ($addr_{upper}$). Consequently, the value of $reg_{Acc}$ at $addr_{upper}$ is used as the upper bound.
When \ctrldata is defined within the same stack frame as $reg_{Acc}$, \textit{base} is used as the buffer's upper bound. Similarly, when $reg_{Acc}$ obtains its definition from another source (e.g., a fixed memory address or as an output from \textit{malloc}), $addr_{upper}$ may not exist or cannot always be determined through this method. In this case, \acron defaults to using \textit{base} as the upper bound.
% ~\oak{dont you want to take a minimum between (base-lowerbound) and (upperbound-lowerbound)? I mean in case of corrupted function pointer, you wouldnt want to upperbound to be the stack top right?} 
%% Adam: My reasoning here is the following: overflow implies the buffer is exceeding its upper bound. Therefore, the lower bound is always where we found reg_Acc to be defined. The upper bound is either (1) the location where the frame is defined (when ctrldata is in another stack frame) or (2) the addr of base itself

Although this strategy could overestimate (i.e., the estimated bounds may include other data objects and the buffer of interest),
narrowing the bounds any further cannot be done without source-code level semantics. For example, the instruction at \texttt{0xe0ea} modifies the stack pointer by ten. However, it cannot be determined through the binary alone whether this is allocating space for a buffer of size ten or for all stack variables totaling ten memory addresses. 
Although this differentiation cannot be made through binary alone, \acron's approach guarantees overflows into \ctrldata are blocked by restricting overflows within the buffer's stack frame.

% After this process, the bounds check to be applied on the buffer can be determined as the offset between $reg_{Acc}^{Init}$ (i.e., \texttt{r11} value) and $base$. 
% Using $MemMap$ and $RegMap$ from the previous phase, Exploit Locator obtains the symbolic expressions of $base$ and $reg_{Acc}^{Init}$ and computes the upper bound of the buffer by evaluating their difference: $bound = reg_{Acc}^{Init} - base$. This value is given to the next module.

\begin{figure}[t]
    \centering
    \includegraphics[width=1\columnwidth]{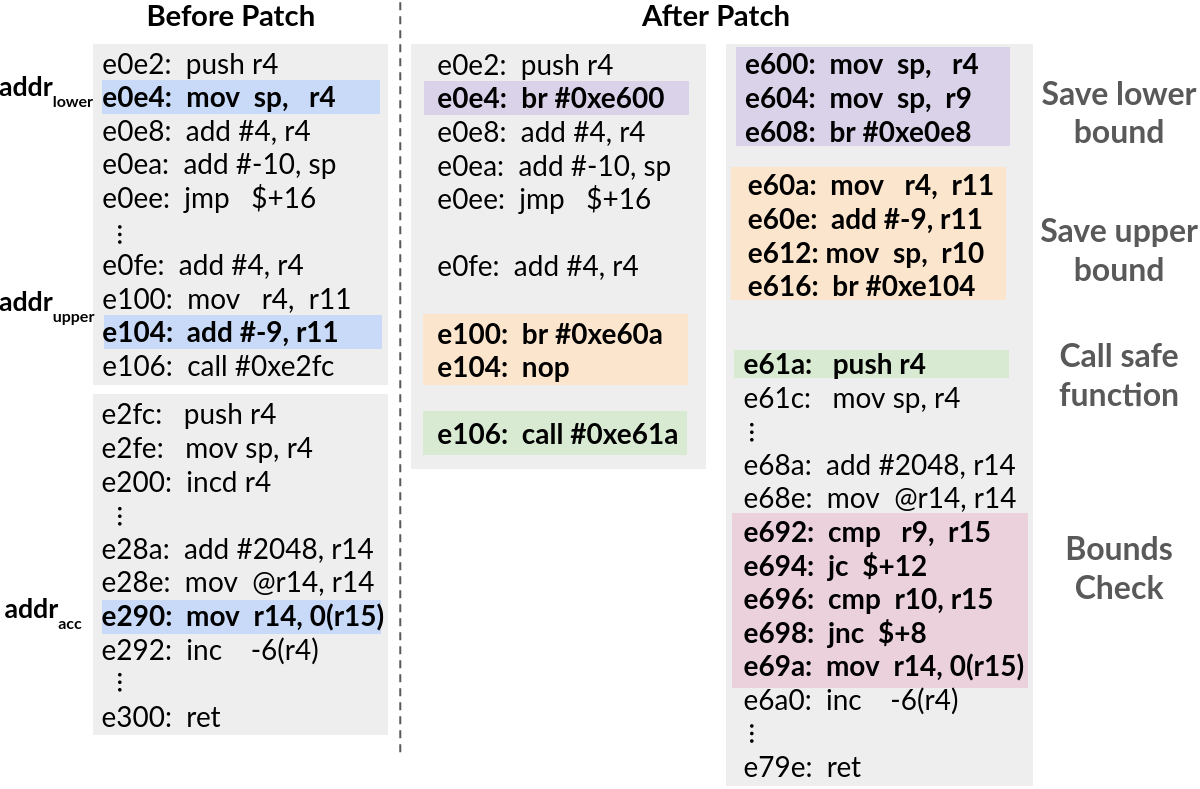}
    \vspace{-2.5em}    
    \caption{Example patch for buffer overflow vulnerabilities.}
    \label{fig:full-ovf-patch}
    \vspace{-2em}
\end{figure}

% \begin{center}
\textbf{T2: Recording the Bounds.}
% \end{center}
After identifying $addr_{lower}$ and $addr_{upper}$, \acron completes \textbf{T1} and has estimated the bounds of the buffer that overflows into \ctrldata. To complete \textbf{T2}, the values of $reg_{Acc}$ must be saved at both locations. This is completed by placing a trampoline both at $addr_{lower}$ and $addr_{upper}$ to preserve their values in a dedicated register. Fig.~\ref{fig:full-ovf-patch} depicts the entire patch for buffer overflow vulnerabilities and shows how these instructions are added for recording the bounds.

In this example, the trampolines and new instructions to save the lower and upper bounds are depicted with the instructions highlighted in purple and orange, respectively. First, the original instructions at $addr_{lower}$ and $addr_{upper}$ are replaced with a trampoline into a new region in the binary. For the upper bound, placing the trampoline alters the alignment, thus the trampoline and a \texttt{nop} are inserted to replace two instructions. Once trampolined into the new region, the instructions that were replaced by the trampoline execute first. Then, $reg_{Acc}$ (\texttt{sp} in this example) is recorded into reserved registers \texttt{r9} and \texttt{r10} to save the lower and upper bounds, respectively. To ensure the bounds cannot be altered, \acron does a pass over \appbin binary to replace any other uses of these registers to ensure they are reserved. After the reserved registers have been updated with $reg_{Acc}$ value, the new instructions trampoline back into the previous code.

% \subsubsection{Generating the Patch}~\\

% \begin{center}
    \textbf{T3: Inserting the Bounds Check}
% \end{center}
Then, Patch Generator proceeds to patch the offending memory instruction by prepending it with instructions that check if the address being accessed is outside the valid boundary.
If it is, the offending instruction is skipped.
For the patch depicted in Fig.~\ref{fig:full-ovf-patch}, the prepended instructions (highlighted in red) correspond to:
\begin{myitemize}
    \item \texttt{cmp r15, r9;} \texttt{jc \$+12;} A comparison between the memory address being accessed (via \texttt{r15}) and the lower bound of the buffer (preserved in \texttt{r9}). If \texttt{r15} is less than \texttt{r9}, it will skip the memory write by jumping to an offset of +12.
    \item \texttt{cmp r15, r10;} \texttt{jnc \$+8;} A similar comparison except to the upper bound of the buffer (preserved in \texttt{r10}). If \texttt{r15} exceeds the upper bound, the memory write will be skipped by jumping to an offset of +8.
\end{myitemize}
The example in Fig.~\ref{fig:full-ovf-patch} uses instructions compiled for TI-MSP430, but the same patch is applied by \acron to ARM-Cortex M33 binaries by using equivalent instructions. For example, instead of using \texttt{jc} and \texttt{jnc}, the patch targeting an ARM-Cortex M33 binary uses \texttt{bcc} and \texttt{bhi}, respectively.

To complete \textbf{T3}, the entire function that contains the exploited memory instruction ($addr_{acc}$) is rewritten to include this bounds check. The modified ``safe'' version of this function is placed into a new region in unused memory. We chose to make a safe copy and do not alter the original function in place to avoid altering all uses of this function (as the vulnerability is specific to this usage and does not necessarily apply to all calls in the code). As such, the final task \textbf{T4} ensures that the safe version of this function is called at the previous \ctrldata-corrupting call site \ctrldata.

% \begin{center}
    \textbf{T4: Calling the Safe Function.}
% \end{center}
To realize \textbf{T4}, Patch Generator rewrites the original \texttt{call} instruction that leads to the function containing the buffer overflow. It changes this \texttt{call} instruction to instead call the patched version. This is depicted in Fig.~\ref{fig:full-ovf-patch} by the instructions highlighted in green.

The \texttt{call} instruction at \texttt{0xe106} previously called to the ``unsafe'' function at \texttt{0xe2fc}. In the patched version, this \texttt{call} instruction now calls \texttt{0xe61a}, the location of the patched version of the function. As depicted, this new function contains bounds-checking from \textbf{T3}.

\subsection{Patch Validator}
\label{subsec:patch_validation}

Before deploying the patched software on \prv, it must be validated to ensure its effectiveness against the previously identified attack. 
%To achieve this, Patch Validator symbolically executes the \slice again but on the now-patched binary. 
%The patch is considered successful if \ctrldata is not overwritten during \slice execution.
% Since the \cflog beyond the range of the slice contains an anomalous/arbitrary path, the patch only pertains to the control flow transfers within the \slice. 
To achieve this, this module first builds an updated CFG of the patched binary. Then, it remaps the destination addresses in \slice to their respective addresses in the updated CFG.
It also inserts new entries into \slice as necessary to account for the patching behaviors (e.g., a trampoline to/from the patch).
% OAK: why do we need to remap the whole \cflog? We only care \slice (a subset of \cflog) right?
%so that any control flow transfers to/from a patched node now reference their new address. 
%With the patched CFG and the remapped \cflog, this module re-executes the same slice of the remapped \cflog.
Patch Validator then performs symbolic DF analysis over the translated \slice, similar to how it is done in Sec.~\ref{subsec:locating_mem_instr}.
When the analysis reaches the end of \slice without \ctrldata being corrupted,
Patch Validator considers this patch effective.
% OAK: say anything about the otherwise case?
% Adam: we can add this based on what we discussed in the meeting:
% OAK: good!
When the corruption still occurs, the attack was likely caused by another vulnerability. In this case, \acron generates a report indicating to \vrf that further manual analysis is required.

\newcommand{\subplotwidth}{4.6cm}

%%%%% change bar oclors 
\newcommand{\ArmUAFcolor}{blue}
\newcommand{\ArmOVFcolor}{orange}
\newcommand{\MspUAFcolor}{gray}
\newcommand{\MspOVcolor}{white}

\newcommand{\bottomRowWidth}{0.195\textwidth}

\begin{figure*}[h]
    \centering    
    {
    \hfill
    \begin{subfigure}{0.19\textwidth} % Adjust width so all subfigures span text width
    \centering
    \begin{tikzpicture}
        \begin{axis}[
            width=\subplotwidth,
            height=4cm,
            ymin=0, ymax=80,
            symbolic x coords={A,B,C,D,E,F,G},
            xtick=data,
            xticklabels={\footnotesize{libbs}, \footnotesize{lcdnum}, \footnotesize{jfdctint}, \footnotesize{fibcall}, \footnotesize{crc\_32}, \footnotesize{cover}, \footnotesize{compress}},
            x tick label style={rotate=45, anchor=north east}, % Rotate labels
            ybar=0.1pt,
            bar width=2pt
        ]
            %% ARM uaf
            \addplot+[ybar, fill=\ArmUAFcolor, draw=black, line width=0.25pt] coordinates {
            (A, 9.8825) (B, 10.2409) (C, 12.6098) (D, 10.0609) (E, 9.8599) (F, 16.6729) (G, 12.1403) };

            %%% ARM OVF
            \addplot+[ybar, fill=\ArmOVFcolor, draw=black, line width=0.25pt] coordinates { (A, 9.4613) (B, 10.6009) (C, 11.8865) (D, 9.8568) (E, 9.7541) (F, 17.2316) (G, 12.0435)  };
            % 0.0205, 0.0379, 0.0181, 0.0343, 0.7136, 0.1653, 0.0205

            %%% MSP UAF
            \addplot+[ybar, fill=\MspUAFcolor, draw=black, line width=0.25pt] coordinates {(A, 66.2868) (B, 66.1901) (C, 70.2832) (D, 66.2781) (E, 66.675) (F, 69.4351) (G, 67.9883) };

            %%% MSP OVF
            \addplot+[ybar, fill=\MspOVcolor, draw=black, line width=0.25pt] coordinates {(A, 66.8693) (B, 65.4758) (C, 69.7086) (D, 65.7042) (E, 66.2129) (F, 69.1135) (G, 67.6122) };
        \end{axis}
    \end{tikzpicture}
    \vspace{-2em}
    \caption{Build CFG}
    \label{fig:build_cfg}
\end{subfigure}
    \hfill
    \begin{subfigure}{0.19\textwidth} % Adjust width so all subfigures span text width
    \centering
    \begin{tikzpicture}
        \begin{axis}[
            width=\subplotwidth,
            height=4cm,
            ymin=0, ymax=40,
            symbolic x coords={A,B,C,D,E,F,G},
            xtick=data,
            xticklabels={\footnotesize{libbs}, \footnotesize{lcdnum}, \footnotesize{jfdctint}, \footnotesize{fibcall}, \footnotesize{crc\_32}, \footnotesize{cover}, \footnotesize{compress}},
            x tick label style={rotate=45, anchor=north east}, % Rotate labels
            ybar=0.1pt,
            bar width=2pt
        ]
            %% ARM uaf
            \addplot+[ybar, fill=\ArmUAFcolor, draw=black, line width=0.25pt] coordinates {
            (A, 1.6319) (B, 2.095) (C, 1.4834) (D, 2.0152) (E, 2.4157) (F, 20.1128) (G, 9.3291) };

            %%% ARM OVF
            \addplot+[ybar, fill=\ArmOVFcolor, draw=black, line width=0.25pt] coordinates { (A, 9.0608) (B, 9.9419) (C, 7.4125) (D, 9.4914) (E, 34.7355) (F, 20.0985) (G, 8.1907)  };
            % 0.0205, 0.0379, 0.0181, 0.0343, 0.7136, 0.1653, 0.0205

            %%% MSP UAF
            \addplot+[ybar, fill=\MspUAFcolor, draw=black, line width=0.25pt] coordinates {(A, 1.1811) (B, 1.5924) (C, 1.4515) (D, 1.7371) (E, 2.0996) (F, 11.049) (G, 1.3857) };

            %%% MSP OVF
            \addplot+[ybar, fill=\MspOVcolor, draw=black, line width=0.25pt] coordinates {(A, 2.142) (B, 2.4191) (C, 2.405) (D, 2.586) (E, 30.8325) (F, 14.209) (G, 3.2328) };
        \end{axis}
    \end{tikzpicture}
    \vspace{-2em}
    \caption{Verify \cflog}
    \label{fig:verify_path}
\end{subfigure}
    \hfill
    \begin{subfigure}{0.19\textwidth} % Adjust width so all subfigures span text width
    \centering
    \begin{tikzpicture}
        \begin{axis}[
            width=\subplotwidth,
            height=4cm,
            ymin=0, ymax=6,
            symbolic x coords={A,B,C,D,E,F,G},
            xtick=data,
            xticklabels={\footnotesize{libbs}, \footnotesize{lcdnum}, \footnotesize{jfdctint}, \footnotesize{fibcall}, \footnotesize{crc\_32}, \footnotesize{cover}, \footnotesize{compress}},
            x tick label style={rotate=45, anchor=north east}, % Rotate labels
            ybar=0.1pt,
            bar width=2pt
        ]
            %% ARM uaf
            \addplot+[ybar, fill=\ArmUAFcolor, draw=black, line width=0.25pt] coordinates {
            (A, 0.2047) (B, 0.3394) (C, 0.4688) (D, 0.1382) (E, 0.7507) (F, 3.5582) (G, 0.3002)};

            %%% ARM OVF
            \addplot+[ybar, fill=\ArmOVFcolor, draw=black, line width=0.25pt] coordinates { (A, 0.0205) (B, 0.0379) (C, 0.0181) (D, 0.0343) (E, 0.7136) (F, 0.1653) (G, 0.0205) };
            % 0.0205, 0.0379, 0.0181, 0.0343, 0.7136, 0.1653, 0.0205

            %%% MSP UAF
            \addplot+[ybar, fill=\MspUAFcolor, draw=black, line width=0.25pt] coordinates { (A, 0.2191) (B, 0.2488) (C, 5.1711) (D, 0.1706) (E, 0.8832) (F, 1.7796) (G, 0.3132) };

            %%% MSP OVF
            \addplot+[ybar, fill=\MspOVcolor, draw=black, line width=0.25pt] coordinates {(A, 0.0208) (B, 0.0264) (C, 0.026) (D, 0.0282) (E, 0.3729) (F, 0.0877) (G, 0.0296) };
        \end{axis}
    \end{tikzpicture}
    \vspace{-2em}
    \caption{Backwards Traversal}
    \label{fig:back_trace}
\end{subfigure}
    \hfill
    \begin{subfigure}{0.19\textwidth} % Adjust width so all subfigures span text width
    \centering
    \begin{tikzpicture}
        \begin{axis}[
            width=4.4cm,
            height=4cm,
            ymin=0, ymax=20000,
            ytick={0, 5000, 10000, 15000, 20000},
            yticklabels={0, 5K, 10K, 15K, 20K},
            symbolic x coords={A,B,C,D,E,F,G},
            scaled y ticks=false,
            xtick=data,
            xticklabels={\footnotesize{libbs}, \footnotesize{lcdnum}, \footnotesize{jfdctint}, \footnotesize{fibcall}, \footnotesize{crc\_32}, \footnotesize{cover}, \footnotesize{compress}},
            x tick label style={rotate=45, anchor=north east}, % Rotate labels
            ybar=0.1pt,
            bar width=2pt
        ]
            %% ARM uaf
            \addplot+[ybar, fill=\ArmUAFcolor, draw=black, line width=0.25pt] coordinates {
            (A, 884.567) (B, 898.9456) (C, 930.466) (D, 872.7318) (E, 875.9085) (F, 945.8547) (G, 916.9845) };

            %%% ARM OVF
            \addplot+[ybar, fill=\ArmOVFcolor, draw=black, line width=0.25pt] coordinates { (A, 6134.0448) (B, 7177.7613) (C, 19525.7302) (D, 4315.9465) (E, 3437.3204) (F, 3441.8717) (G, 6930.996)};
            % 0.0205, 0.0379, 0.0181, 0.0343, 0.7136, 0.1653, 0.0205

            %%% MSP UAF
            \addplot+[ybar, fill=\MspUAFcolor, draw=black, line width=0.25pt] coordinates { (A, 2679.8927) (B, 2650.0946) (C, 2627.8099) (D, 2624.5653) (E, 2679.3883) (F, 2682.7707) (G, 2753.4829) };

            %%% MSP OVF
            \addplot+[ybar, fill=\MspOVcolor, draw=black, line width=0.25pt] coordinates {(A, 2169.2883) (B, 2421.8644) (C, 1698.3695) (D, 1662.7329) (E, 2643.1913) (F, 2167.557) (G, 2677.7727)};
        \end{axis}
    \end{tikzpicture}
    \vspace{-2em}
    \caption{Symb. DF Analysis}
    \label{fig:symb_df}
\end{subfigure}
    \hfill
    \begin{subfigure}{0.19\textwidth} % Adjust width so all subfigures span text width
    \centering
    \begin{tikzpicture}
        \begin{axis}[
            width=\subplotwidth,
            height=4cm,
            ymin=0, ymax=10,
            symbolic x coords={A,B,C,D,E,F,G},
            xtick=data,
            xticklabels={\footnotesize{libbs}, \footnotesize{lcdnum}, \footnotesize{jfdctint}, \footnotesize{fibcall}, \footnotesize{crc\_32}, \footnotesize{cover}, \footnotesize{compress}},
            x tick label style={rotate=45, anchor=north east}, % Rotate labels
            ybar=0.1pt,
            bar width=2pt
        ]
            %% ARM uaf
            \addplot+[ybar, fill=\ArmUAFcolor, draw=black, line width=0.25pt] coordinates {
            (A, 0.5718) (B, 0.5929) (C, 0.5646) (D, 0.5799) (E, 0.5826) (F, 0.5815) (G, 0.5787) };

            %%% ARM OVF
            \addplot+[ybar, fill=\ArmOVFcolor, draw=black, line width=0.25pt] coordinates { (A, 9.3093) (B, 8.7548) (C, 8.6592) (D, 9.1652) (E, 8.5663) (F, 8.6368) (G, 8.6348) };
            % 0.0205, 0.0379, 0.0181, 0.0343, 0.7136, 0.1653, 0.0205

            %%% MSP UAF
            \addplot+[ybar, fill=\MspUAFcolor, draw=black, line width=0.25pt] coordinates { (A, 0.1597) (B, 0.1573) (C, 0.1573) (D, 0.1615) (E, 0.1604) (F, 0.1548) (G, 0.1577) };

            %%% MSP OVF
            \addplot+[ybar, fill=\MspOVcolor, draw=black, line width=0.25pt] coordinates {(A, 2.5195) (B, 2.5471) (C, 2.3024) (D, 2.3672) (E, 2.4953) (F, 2.886) (G, 2.4876) };
        \end{axis}
    \end{tikzpicture}
    \vspace{-2em}
    \caption{Patch Generation}
    \label{fig:patch_gen}
\end{subfigure}
    \hfill
    \begin{subfigure}{\bottomRowWidth} % Adjust width so all subfigures span text width
    \centering
    \begin{tikzpicture}
        \begin{axis}[
            width=\subplotwidth,
            height=4cm,
            ymin=0, ymax=60,
            symbolic x coords={A,B,C,D,E,F,G},
            xtick=data,
            xticklabels={\footnotesize{libbs}, \footnotesize{lcdnum}, \footnotesize{jfdctint}, \footnotesize{fibcall}, \footnotesize{crc\_32}, \footnotesize{cover}, \footnotesize{compress}},
            x tick label style={rotate=45, anchor=north east}, % Rotate labels
            ybar=0.1pt,
            bar width=2pt
        ]
            %% ARM uaf
            \addplot+[ybar, fill=\ArmUAFcolor, draw=black, line width=0.25pt] coordinates {
             (A, 8.1446) (B, 8.3169) (C, 8.8741) (D, 8.1041) (E, 8.1403) (F, 9.4979) (G, 8.6104) };

            %%% ARM OVF
            \addplot+[ybar, fill=\ArmOVFcolor, draw=black, line width=0.25pt] coordinates { (A, 48.402) (B, 47.1057) (C, 47.6593) (D, 48.7983) (E, 46.2992) (F, 47.0869) (G, 45.999) };
            % 0.0205, 0.0379, 0.0181, 0.0343, 0.7136, 0.1653, 0.0205

            %%% MSP UAF
            \addplot+[ybar, fill=\MspUAFcolor, draw=black, line width=0.25pt] coordinates { (A, 12.2985) (B, 12.2509) (C, 13.022) (D, 12.2396) (E, 12.49) (F, 12.5855) (G, 12.3575) };

            %%% MSP OVF
            \addplot+[ybar, fill=\MspOVcolor, draw=black, line width=0.25pt] coordinates {(A, 49.2765) (B, 48.7976) (C, 48.3755) (D, 50.0747) (E, 50.3332) (F, 52.3137) (G, 48.9058) };
        \end{axis}
    \end{tikzpicture}
    \vspace{-2em}
    \caption{Update ELF}
    \label{fig:update_elf}
\end{subfigure}
    % \hfill
    \begin{subfigure}{\bottomRowWidth} % Adjust width so all subfigures span text width
    \centering
    \begin{tikzpicture}
        \begin{axis}[
            width=\subplotwidth,
            height=4cm,
            ymin=0, ymax=3,
            symbolic x coords={A,B,C,D,E,F,G},
            xtick=data,
            xticklabels={\footnotesize{libbs}, \footnotesize{lcdnum}, \footnotesize{jfdctint}, \footnotesize{fibcall}, \footnotesize{crc\_32}, \footnotesize{cover}, \footnotesize{compress}},
            x tick label style={rotate=45, anchor=north east}, % Rotate labels
            ybar=0.1pt,
            bar width=2pt
        ]
            %% ARM uaf
            \addplot+[ybar, fill=\ArmUAFcolor, draw=black, line width=0.25pt] coordinates {
            (A, 0.3247) (B, 0.3325) (C, 0.3203) (D, 0.3295) (E, 0.3341) (F, 1.9137) (G, 1.8577) };

            %%% ARM OVF
            \addplot+[ybar, fill=\ArmOVFcolor, draw=black, line width=0.25pt] coordinates { (A, 2.6113) (B, 2.4902) (C, 2.4998) (D, 2.8031) (E, 2.2958) (F, 2.3918) (G, 2.6113) };
            % 0.0205, 0.0379, 0.0181, 0.0343, 0.7136, 0.1653, 0.0205

            %%% MSP UAF
            \addplot+[ybar, fill=\MspUAFcolor, draw=black, line width=0.25pt] coordinates { (A, 0.1839) (B, 0.2037) (C, 0.2125) (D, 0.2144) (E, 0.2526) (F, 0.7858) (G, 0.1972) };

            %%% MSP OVF
            \addplot+[ybar, fill=\MspOVcolor, draw=black, line width=0.25pt] coordinates {(A, 1.0971) (B, 1.2251) (C, 0.8421) (D, 0.8251) (E, 2.8966) (F, 1.9175) (G, 0.8109) };
        \end{axis}
    \end{tikzpicture}
    \vspace{-1em}
    \caption{Translate \cflog}
    \label{fig:translate_cflog}
\end{subfigure}
    % \hfill
    \begin{subfigure}{\bottomRowWidth} % Adjust width so all subfigures span text width
    \centering
    \begin{tikzpicture}
        \begin{axis}[
            width=4.4cm,
            height=4cm,
            ymin=0, ymax=100,
            symbolic x coords={A,B,C,D,E,F,G},
            xtick=data,
            xticklabels={\footnotesize{libbs}, \footnotesize{lcdnum}, \footnotesize{jfdctint}, \footnotesize{fibcall}, \footnotesize{crc\_32}, \footnotesize{cover}, \footnotesize{compress}},
            x tick label style={rotate=45, anchor=north east}, % Rotate labels
            ybar=0.1pt,
            bar width=2pt
        ]
            %% ARM uaf
            \addplot+[ybar, fill=\ArmUAFcolor, draw=black, line width=0.25pt] coordinates {
            (A, 9.7041) (B, 10.2417) (C, 12.1532) (D, 9.62) (E, 9.4678) (F, 15.8887) (G, 11.6786)  };

            %%% ARM OVF
            \addplot+[ybar, fill=\ArmOVFcolor, draw=black, line width=0.25pt] coordinates { (A, 9.1616) (B, 9.116) (C, 11.7527) (D, 8.2396) (E, 8.2603) (F, 16.0668) (G, 11.159) };
            % 0.0205, 0.0379, 0.0181, 0.0343, 0.7136, 0.1653, 0.0205

            %%% MSP UAF
            \addplot+[ybar, fill=\MspUAFcolor, draw=black, line width=0.25pt] coordinates {  (A, 68.0698) (B, 68.2021) (C, 72.7587) (D, 67.8639) (E, 69.2667) (F, 71.2734) (G, 70.2204)  };

            %%% MSP OVF
            \addplot+[ybar, fill=\MspOVcolor, draw=black, line width=0.25pt] coordinates {(A, 68.6794) (B, 68.5184) (C, 72.4066) (D, 68.7065) (E, 93.772) (F, 96.952) (G, 95.4205) };
        \end{axis}
    \end{tikzpicture}
    \vspace{-2em}
    \caption{Build Patched CFG}
\end{subfigure}
    % \hfill
    \begin{subfigure}{\bottomRowWidth} % Adjust width so all subfigures span text width
    \centering
    \begin{tikzpicture}
        \begin{axis}[
            width=\subplotwidth,
            height=4cm,
            ymin=0, ymax=16000,
            ytick={0, 4000, 8000, 12000, 16000},
            yticklabels={0, 4K, 8K, 12K, 16K},
            scaled y ticks=false,
            symbolic x coords={A,B,C,D,E,F,G},
            xtick=data,
            xticklabels={\footnotesize{libbs}, \footnotesize{lcdnum}, \footnotesize{jfdctint}, \footnotesize{fibcall}, \footnotesize{crc\_32}, \footnotesize{cover}, \footnotesize{compress}},
            x tick label style={rotate=45, anchor=north east}, % Rotate labels
            ybar=0.1pt,
            bar width=2pt
        ]
            %% ARM uaf
            \addplot+[ybar, fill=\ArmUAFcolor, draw=black, line width=0.25pt] coordinates {
            (A, 833.9449) (B, 854.4547) (C, 825.0547) (D, 824.8444) (E, 828.6351) (F, 852.0378) (G, 872.1885) };

            %%% ARM OVF
            \addplot+[ybar, fill=\ArmOVFcolor, draw=black, line width=0.25pt] coordinates { (A, 4822.3276) (B, 5544.5354) (C, 14286.1473) (D, 3505.0592) (E, 2885.5163) (F, 2906.5243) (G, 5543.2039) };
            % 0.0205, 0.0379, 0.0181, 0.0343, 0.7136, 0.1653, 0.0205

            %%% MSP UAF
            \addplot+[ybar, fill=\MspUAFcolor, draw=black, line width=0.25pt] coordinates { (A, 2990.0045) (B, 2898.8526) (C, 2929.881) (D, 3042.6285) (E, 2916.2127) (F, 2925.0125) (G, 3036.079)  };

            %%% MSP OVF
            \addplot+[ybar, fill=\MspOVcolor, draw=black, line width=0.25pt] coordinates {(A, 1278.3783) (B, 1379.2985) (C, 1034.7473) (D, 1026.5723) (E, 1479.5068) (F, 1244.827) (G, 1500.5279) };
        \end{axis}
    \end{tikzpicture}
    \vspace{-2em}
    \caption{Symb. Patch Validation}
    \label{fig:re_exec_slice}
\end{subfigure}
    \hfill
    }
    \vspace{-0.5em}
    \begin{subfigure}{\textwidth} % Adjust width so all subfigures span text width
    \centering
    % \vspace{-2pt}
    \begin{tikzpicture}
        %% Why are the distances between nodes non-constant but they appear constant in the pdf? Don't ask

        %% You might also be wondering about the units of tikzpicture's coordinate system.
        % again, don't ask
        \draw[black, line width=0.2pt] (-4.5,0.3) rectangle (13,0.3);
    
        \node[text width=3cm] at (0,0) {\textbf{Legend:}};
        
        %% ARM uaf \ArmUAFcolor
        \node[text width=3cm] at (1.5,0) {ARM-UAF: };
        \fill[\ArmUAFcolor] (1.7,-0.2) rectangle (2.2,0.2);
        
        %%% ARM OVF \ArmOVFcolor
        \node[text width=3cm] at (4,0) {ARM-OVF: };
        \fill[\ArmOVFcolor] (4.2,-0.2) rectangle (4.7,0.2);

        %%% MSP UAF \MspUAFcolor
        \node[text width=3cm] at (6.5,0) {MSP-UAF: };
        \fill[\MspUAFcolor] (6.7,-0.2) rectangle (7.2,0.2);
        % \draw[black, line width=1pt] (6.7,-0.2) rectangle (7.2,0.2);            
        
        %%% MSP OVF \MspOVcolor
        \node[text width=3cm] at (9,0) {MSP-OVF: };
        \fill[\MspOVcolor] (9.2,-0.2) rectangle (9.7,0.2);
        \draw[black, line width=1pt] (9.2,-0.2) rectangle (9.7,0.2);
    \end{tikzpicture}
    \vspace{-1.5em}
\end{subfigure}
    \vspace{-0.5em}
    \caption{Run-time of \acron submodules (ms) to analyze evidence of buffer overflow (OVF) and use-after-free (UAF) vulnerabilities on two MCUs and \CFA architectures: hardware-based \CFA atop MSP430~\cite{acfa} and TEE-based \CFA atop ARM Cortex-M33~\cite{traces}.}
    \vspace{-1em}
    \label{fig:module_timing}
\end{figure*}

\section{Implementation \& Evaluation}\label{sec:impl}

\acron was implemented as a combination of Bash and Python and evaluated on an Ubuntu 18.04 machine with an Intel Core i7-4790 CPU at 3.60GHz and 32GB RAM. BEEBs benchmark applications for embedded platforms~\cite{beebs} were used for the evaluation. They were compiled for MSP430 and ARM Cortex-M33 and executed atop unmodified \CFA architectures, ACFA~\cite{acfa} and TRACES~\cite{traces}, to generate a dataset of executables and corresponding \cflog-s.
The MSP430 prototype was configured with 8 KB of application program memory, and the ARM Cortex-M33 prototype was configured with 256 KB of application program memory.
We believe that \acron concepts generalize to several CFA architectures and use ACFA and TRACES as case studies due to their open-source availability.

To obtain CFGs from executables, \acron first runs \textit{objdump} %~\cite{msp430_objdump,arm_objdump} % OAK: its okay to not cite it. I think everyone knows what objdump is.
over the executables to decode the assembly instructions. 
Then, it builds the CFG from this disassembled file. \acron uses SymPy~\cite{sympy} for evaluating symbolic expressions and Keystone assembler~\cite{keystone-engine} for generating ARM Cortex-M instructions. Since no MSP430 assembler exists in Python packages, we build upon and modify MSProbe~\cite{msprobe} for MSP430 instructions. Finally, \acron uses the Python library ELFtools~\cite{pyelftools} to generate the final patched executable. 
\acron proof of concept prototype is publicly available at~\cite{repo}.

\subsection{Run-times of \acron Sub-Modules}\label{subsec:runtime_eval}
We evaluate the run-time of each \acron sub-module after a malicious path was detected.
Given the lack of reproducible exploits and incomplete CVEs for MCUs software (e.g., missing source codes, broken links, etc.)~\cite{tan2024sok}, we craft control flow attacks that exploit buffer overflow and use-after-free vulnerabilities, following a similar methodology to recent related work~\cite{rage}. We then insert them into the example benchmark applications~\cite{beebs} to closely mimic existing CVEs related to control flow hijacking and ROP in MCUs~\cite{cve_2021_0920,cve_2017_14201,cve_2019_16127,cve_2020_10019,cve_2020_10023,cve_2021_35395,cve_2022_34835} to the best of our ability.
For consistency across applications, we insert the vulnerability into the last function called by the main. 
The vulnerability is used to overwrite \ctrldata to return/jump to the main function, causing an infinite loop. We execute the vulnerable programs on two unmodified \CFA architectures -- a TEE-based approach built atop ARM Cortex-M33 (TRACES~\cite{traces}) and a custom hardware-based approach built atop TI MSP430 (ACFA~\cite{acfa}) -- to collect \cflog-s of the attack, which are used by \acron with the modified application binary (\appbin) for further analysis. Fig.~\ref{fig:module_timing} shows the run-time of each \acron sub-module during the analysis.

% \subsubsection{Static Analysis}\label{sec:runtime_static_analysis}
\textbf{Static Analysis.}
The time required to build the CFG depends on the number of branch instructions. Building the CFG for MSP430 binaries requires more run-time, albeit smaller executables, as shown in Fig.~\ref{fig:build_cfg}. This is due to jump destinations in disassembled MSP430 binaries being decoded with offsets, not full addresses~\cite{msp430_objdump}. Therefore, MSP430 binaries require additional processing steps to determine addresses from the offsets, unlike ARM binaries, which disassemble into full addresses~\cite{arm_objdump}. Generating the CFG from MSP430 binaries required $\approx$66.27 to 70.21ms, whereas ARM binaries required $\approx$9.46 to 17.23ms, depending on the application.

% \subsubsection{Path Verifier}
\textbf{Path Verifier.}
Fig.~\ref{fig:verify_path} presents the time to verify \cflog. The verification time depends on the number of branch instructions executed by the program before the corrupted \ctrldata is used. In most cases, the crafted OVF generates a larger \cflog since it introduces more control flow transfers before corrupting \ctrldata compared to the crafted UAF examples. Additionally, \cflog-s of ARM executables (generated by TRACES~\cite{traces}) require longer run-time to verify than \cflog-s of MSP430 executables (generated by ACFA~\cite{acfa}). This is because \cflog-s from TRACES do not contain any static branch instruction destinations, whereas \cflog-s from ACFA contain all control flow transfers. Consequently, TRACES \cflog-s require more time to verify since more processing is required by \vrf to account for the missing entries, whereas ACFA \cflog-s can be naively followed by \vrf without additional effort. The detection time through \cflog verification for the evaluated examples ranged from $\approx$1.14 to 34.74ms.

% \subsubsection{Exploit Locator}
\textbf{Exploit Locator.}
Run-time for Backwards Traversal and Symbolic DF Analysis and are shown in Fig.~\ref{fig:back_trace} and Fig.~\ref{fig:symb_df}, respectively. The Symbolic DF Analysis incurs the highest run-time from 0.868s to 19.5s. For the OVF example in ARM, \slice of the \textit{jfdctint} program covers the highest number of instructions (7750 in total), thus having the longest run-time. In MSP430, the \textit{aha-compress} program (abbreviated \textit{compress}) has the same characteristic. 

Since tracking of memory and registers is not required for Backwards Traversal, its run-time is more efficient in comparison ($\approx$0.026 to 5.31ms). For a similar reason as described above, Backwards Traversal of UAF in \textit{jfdctint} requires the longest run-time.
% The run-time of Backwards Traversal for \textit{cover} also stands out. 
The \textit{cover} application is designed to test the performance while traversing every path in multiple large switch statements.
Thus, the run-time slowdown arises due to tracking definitions across the memory stores/loads of each switch case. This effect is especially apparent in ARM binaries, which implement more memory operations for switch statements than MSP430. Overall, the run-time of the Exploit Locator is mainly dominated by the Symbolic DF Analysis.

%\subsubsection{Patch Generator}
\textbf{Patch Generator.}
%
% Patches are generated by first inserting/updating instructions pertaining to the vulnerability location. Then, \appbin is updated by adding a new dedicated section with the generated instructions. Run-time for these steps are shown in Fig.~\ref{fig:patch_gen} and Fig.~\ref{fig:update_elf}, respectively.
% %
%
Overall, the UAF patch is produced faster than the OVF patch due to its simplicity, as shown in Fig.~\ref{fig:patch_gen} and Fig.~\ref{fig:update_elf}. 
Generating ARM Cortex-M33 patches requires more run-time due to its larger instruction set, differences in the compilation process (i.e., functions in ARM Cortex-M tend to include more loads/stores), and differences in the compiler tools.

% \subsubsection{Patch Validator}
\textbf{Patch Validator.}
%
% Run-times of each step to verify MSP430 and ARM Cortex-M33 patches are shown in Fig.~\ref{fig:translate_cflog}, Fig.~\ref{fig:patch_cfg}, and Fig.~\ref{fig:re_exec_slice}.
%
The time to translate the \cflog, shown in Fig.~\ref{fig:translate_cflog}, depends on the number of new branch instructions introduced as a part of the patch. Therefore, translating \cflog for the UAF test cases requires negligible run-time (e.g., $<$1.0ms) since it introduces no new branch instructions, whereas the OVF test case incurs time due to new branches introduces with patch.

% \begin{table}[t]
% \centering
% \caption{Binary size comparison (bytes) for OVF patches}
% \vspace{-0.8em}
% \label{tab:bin_size_ovf}
% \resizebox{\columnwidth}{!}{
% \begin{tabular}{|c|c|c|c|c|c|c|}
% \hline
% \textbf{MCU Platform} & \multicolumn{3}{c|}{\textbf{MSP430}} & \multicolumn{3}{c|}{\textbf{ARM Cortex-M33}} \\
% \hline
% \textbf{Application} & \textbf{Original} & \textbf{Patched} & \textbf{Increase (\%)} & \textbf{Original} & \textbf{Patched} & \textbf{Increase (\%)} \\
% \hline
% \textit{libbs}        & 808    & 1072   & 32.7\% & 5212 & 5404 & 3.68\% \\
% \textit{lcdnum}       & 872    & 1136    & 30.3\% & 5692 & 5880 & 3.3\% \\
% \textit{jfdctint}     & 5764   & 6028   & 4.58\% & 7292 & 7480 & 2.58\% \\
% \textit{fibcall}      & 792    & 1056   & 33.3\% & 5308 & 5496 & 3.54\% \\
% \textit{crc32}        & 1236   & 1500   & 21.4\% & 5228 & 5416 & 3.6\% \\
% \textit{cover}        & 3516   & 3780   & 7.51\% & 10268 & 10460 & 1.87\%  \\
% \textit{compress} & 2940   & 3208   & 9.12\% & 6796 & 6988 & 2.83\% \\
% \hline  
% \end{tabular}
% }
% \end{table}

The time to generate the patched CFG depends on the architecture. For the UAF case, the run-time is approximately the same since the patch effectively removed one CFG node. The OVF case requires more time to rewrite additional nodes. This difference is more pronounced for MSP430 due to decoding of jump instructions, as previously described for Static Analysis.

Like the Exploit Locator, the run-time of the Patch Validator is dominated by the Symbolic DF Analysis (from 0.82s to 14.3s in Fig.~\ref{fig:re_exec_slice}). 
Nonetheless, this analysis is performed faster in the Patch Validator because the patch causes the previously vulnerable node to exit more quickly.
The bottleneck of the Symbolic DF Analysis comes from emulating memory access using \textit{MemMap}. \acron also requires an additional check for each memory write to determine whether the write corrupts \ctrldata (recall Sec.~\ref{subsec:locating_mem_instr}).
Future work could improve \acron's modeling of memory accesses.

%\vspace{-3mm}
\subsection{Patch Sizes}
% Table~\ref{tab:bin_size_ovf} shows the increase in binary size (for MSP430 and ARM Cortex-M33) before and after the OVF patch for each application. The baseline (i.e., pre-patched) binary size is larger in ARM Cortex-M33 than in MSP430 due to several reasons. 
% First, compared to the 16-bit MSP430 ISA, ARM Cortex-M is 32-bit and thus requires a larger instruction size.
%thus, most of the application is implemented with 32-bit instructions. 
% Second, the ARM Cortex-M33 binaries require additional instrumentation by the underlying \CFA architecture TRACES~\cite{traces} for secure generation of run-time evidence.
We compare program memory size before and after the patch is installed. As a baseline, the MSP430 binaries consumed from 792 to 5764 bytes (9.90\% to 72.1\% of the program memory) 
%% Adam -- Done!
% \oak{Can you put the total MSP430 program memory here in parentheses? It doesnt seem to hurt to show the total number here and people might complain if we report everything in \% (I did in some papers in the past)} 
and the ARM Cortex-M33 binaries consumed from 5.3 KB to 10.3 KB (0.31\% to 2.25\% of the program memory).
% \oak{same here}.
The difference in baseline is due to the MSP430 prototype being much more resource-constrained than the ARM Cortex-M33 prototype (see Sec.~\ref{sec:impl}).

After the OVF patch, the MSP430 binaries consumed 1056 to 6028 bytes (13.2\% to 75.4\% of the program memory) and ARM Cortex-M33 binaries consumed 5.4 KB to 10.4 KB (from 0.41\% to 2.35\% of the program memory). This represents an average increase of $\approx$264.5 bytes ($\approx$3.31\%) 
% \oak{if reporting the whole number looks good, we should also report it here. I remember its like 300 bytes right?}
%% adam - yes, done!
and $\approx$189.7 bytes ($\approx$0.10\%), respectively. 

Since the UAF patch simply replaces a \texttt{call} instruction in place with an equivalent number of \texttt{nop}-s, it incurs no increase in the binary size. However, it incurs additional data memory overhead at run-time due to the missing \texttt{free}.
The data memory overhead depends on the size of the object that is no longer freed.

\begin{comment}
%%%% i have this here just to have everything, but its probably not very reasonable to add it in

\begin{table}[t]
\centering
\caption{Binary size comparison (bytes) for use-after-free patch}
\label{tab:bin_size_uaf}
\resizebox{\columnwidth}{!}{
\begin{tabular}{|c|c|c|c|c|c|c|}
\hline
\textbf{MCU Platform} & \multicolumn{3}{c|}{\textbf{MSP430}} & \multicolumn{3}{c|}{\textbf{ARM Cortex-M33}} \\
\hline
\textbf{Application} & \textbf{Original} & \textbf{Patched} & \textbf{Increase (\%)} & \textbf{Original} & \textbf{Patched} & \textbf{Increase (\%)} \\
\hline
\textit{libbs}        & 1372    & 1372   & 0.00\% & 5948 & 5948 & 0.00\% \\
\textit{lcdnum}       & 1432    & 1432    & 0.00\% & 6172 & 6172 & 0.00\% \\
\textit{jfdctint}     & 6328   & 6328   & 0.00\% & 8012 & 8012 & 0.00\% \\
\textit{fibcall}      & 1364    & 1364   & 0.00\% & 6028 & 6028 & 0.00\% \\
\textit{crc32}        & 1816   & 1816   & 0.00\% & 5964 & 5964 & 0.00\% \\
\textit{cover}        & 4088   & 4088   & 0.00\% & 10972 & 10972 & 0.00\%  \\
\textit{compress} & 3512   & 3512   & 0.00\% & 7516 & 7516 & 0.00\% \\
\hline  
\end{tabular}
}
\end{table}
\end{comment}
% \begin{figure}[t]
% \centering
% \hfill
% \includegraphics[width=0.8\columnwidth]{figs/binary-size-arm.png}
% \hfill
% %\vspace{0.5em}
% \hfill
% \includegraphics[width=0.8\columnwidth]{figs/binary-size-msp430.png}
% \hfill
% \caption{run-time of \acron submodules}
% % %\vspace{-1em}
% \label{fig:bin_size}
% \end{figure}

\section{Related Work}\label{sec:rw}

%%% old
% \textbf{Control Flow Attestation.} C-FLAT~\cite{cflat} proposed \CFA by
% instrumenting an application binary at branch instructions with calls that trap execution in an ARM TrustZone-protected ``Secure World'', where the branch destination is added to a hash chain representing the current path in a unique digest. 
% Many techniques follow C-FLAT by combining instrumentation and TEE support to improve \CFA~\cite{enola,scarr,recfa,oat,ari,traces,blast} or by employing custom hardware extensions to detect/log control flow transfers without requiring instrumentation/TEE support~\cite{lofat,litehax,atrium,acfa}.
%%
%
% The majority of existing work in this area focuses on the correctness and security of \prv implementation to generate run-time evidence~\cite{cflat,scarr,recfa,oat,ari,lofat,litehax,atrium,tinycfa,acfa,blast,iscflat,enola}. They often mention that with authentic evidence, \vrf can detect the validity of the reported path by utilizing data structures computed prior to run-time (e.g., CFG+SS~\cite{scarr,recfa,oat,ari,acfa,tinycfa} or hash-sets~\cite{lofat,atrium,cflat,blast}) or by emulating execution~\cite{litehax,dialed}. 
% As demonstrated by this work, all techniques that create hash-based/hybrid evidence (\textbf{E1} and \textbf{E3} from Sec.~\ref{sec:classify}) cannot infer the precise malicious execution path on \prv and are not suitable for conducting root cause analysis.

\edit{\textbf{Control Flow Attestation.} 
Current \CFA approaches generate execution path evidence combining binary instrumentation with TEE support~\cite{enola,scarr,recfa,oat,ari,traces,blast} or using custom hardware to detect/log control flow transfers~\cite{lofat,litehax,atrium,acfa}.
Most \CFA literature focuses on securely generating correct evidence on a compromised \prv, mentioning that this evidence can be used with CFG+SS~\cite{scarr,recfa,oat,ari,acfa,tinycfa}, hash-sets~\cite{lofat,atrium,cflat,blast}, or execution emulation~\cite{litehax,dialed} to perform verification. 
As discussed in this paper, techniques that create hash-based/hybrid evidence (\textbf{E1} and \textbf{E3} from Sec.~\ref{sec:classify}) cannot always infer the precise malicious execution paths on \prv and are not suitable for root cause analysis.}

To our knowledge, only two prior efforts~\cite{debes2023zekra,rage} propose approaches for \CFA verification.
\edit{ZEKRA~\cite{debes2023zekra} proposes a \CFA verification in zero-knowledge proofs to ensure that an untrusted \vrf does not learn secrets from \cflog.}
%
% RAGE~\cite{rage} presents a path verification technique that does not require constructing a complete CFG. 
% RAGE trains a Graph Neural Network (GNN) on a set of known \cflog-s and validates new \cflog-s using the GNN. 
\edit{RAGE~\cite{rage} presents a method for path verification by training a Graph Neural Network (GNN) on \cflog-s, removing the need for a CFG.
Importantly, neither ZEKRA nor RAGE supports root cause analysis from received evidence, which is \acron's main goal.
Although \acron requires constructing a partial CFG (unlike RAGE), we show this is feasible for MCU applications; see Fig.~\ref{fig:build_cfg}.
Ammar et al.~\cite{sok_cfa_cfi} provide a systematization of run-time defenses and discuss \CFA verification as an open challenge. Unlike their discussion on the challenges in verifying \CFA evidence, our work analyzes \CFA evidence formats and their suitability for advanced analysis in depth. We also show how \vrf can use \textbf{E2} evidence to pinpoint and remediate root causes concretely.}

{
\color{black}

% \subsection{Root Cause Analysis}
\textbf{Root Cause Analysis.} 
 Root Cause Analysis frameworks
~\cite{yagemann2021arcus,liu2022seeker,park2024benzene,zeng2022palantir,zhou2022ncscope,yagemann2021automated} %% shortened
automate the process for identifying sources of software bugs or unexpected behavior. 
% Two closely related Root Cause Analysis frameworks are ARCUS and BENZENE.
%
ARCUS~\cite{yagemann2021arcus} collects an Intel PT trace and an initial memory snapshot, which are used by an ARCUS server after a run-time violation to identify root cause vulnerabilities during execution emulation and to generate a report.
Due to the cost of storing and transmitting large memory snapshots, ARCUS and similar proposals~\cite{yagemann2021automated,zeng2022palantir,liu2022seeker,zhou2022ncscope} are unlikely to integrate feasibly with \CFA on low-end MCUs.
% (a common platform for \CFA~\cite{sok_cfa_cfi}).
% To provide root cause analysis suitable with \CFA evidence, \acron operates without a snapshot of initial memory (which is not available from \slice).
%
BENZENE~\cite{park2024benzene} starts with the application binary and crash-inducing inputs, and it constructs a data-flow graph (DFG) to backtrack from crashes to track state conditions that lead to crashes and non-crashes.
It then analyzes these conditions to generate likely root cause data-flow events.
Unlike ACRUS and BENZENE, \acron performs root cause analysis without large memory snapshots, crash inputs, or DFGs.
Instead, \acron combines CFG traversal with symbolic data-flow analysis to identify exploited instructions, generate a binary patch for the identified root cause, and verify its effectiveness. 
% BENZENE~\cite{park2024benzene} performs its analysis using the application binary and crash-inducing inputs. It uses a data-flow graph (DFG) to track all influenced values and mutate states to track when the program terminated successfully or crashed again. Similarly to \acron, BENZENE starts from the crash site and performs a backtracking process.
% It then analyzes the crash/non-crash conditions and generates a list of likely root cause data-flow events.
% Unlike BENZENE, \acron does not require data inputs and does not require constructing a DFG. Instead, \acron combines CFG traversal with symbolic data-flow analysis to determine exploited memory instructions without requiring data inputs.
% Unlike BENZENE, \acron generates and verifies a binary patch for the identified root cause. 

% Include related work on binary patching and discussion of deployment challenges in related work that are also apparent in SABRE.

\textbf{Binary Rewriting and Patching.} Automated binary rewriting ~\cite{bin_rewrite_survey} can be used for pre-deployment security measures~\cite{cots_cfi,ccfir,bincfi,reins_bin_memsafe,fi_safe_bin} and for post-deployment patch generation~\cite{osspatcher,alice,embroidery,icspatch,vulmet,egalito,e9patch}.  
Like prior work in this space~\cite{osspatcher,vulmet,alice,embroidery}, \acron patch instructions redirect execution to a secure implementation. 
However, \acron identifies the target through root-cause analysis instead of function matching or other binary analysis methods.
%
% In run-time auditing, \vrf may use binary rewriting for pre-deployment binary instrumentation (as discussed in Sec.~\ref{sec:cfa}) and for post-deployment remediation after \acron analysis.
Real-world deployments assume the target device has a secure software update module to install the rewritten binary. For \acron, no additional assumptions are required since this could be implemented as the remediation function within the run-time auditing workflow~\cite{acfa,traces}.
Deployment challenges, like vendor-specific limits placed on software updates or flash endurance levels, would be accounted for in implementing the remediation function, and thus they are outside the scope of prior works in binary rewriting and \acron.

% \begin{itemize}
%     \item OSSPatcher~\cite{osspatcher} automated function-level binary patches for mobile devices, uses publicly available source-code patches.
%     \item ALICE~\cite{alice} detects and patches weak and broken cryptographic implementations from binaries.
%     \item E9Patch~\cite{e9patch} binary rewriting without knowledge of control flow targets
%     \item Egalito~\cite{Egalito} rewrites binaries without needing layout information
%     \item Binary rewriting for CFI~\cite{cots_cfi,ccfir}
%     \item Embroider~\cite{embroidery} binary patching by matching binaries to source of known vulnerabilities
%     \item ICSPatch~\cite{icspatch} for PLC devices: captures PLC snapshot, rehosts with angr to search dependency graph, generates a hotpatch
%     \item VulMet~\cite{vulmet} for android kernels based on released patches, it determines best insertion location, then calculates a semantically equivalent hot patch.
%     \item binary-rewriting for other preventative security~\cite{reins_bin_memsafe,fi_safe_bin} 
% \end{itemize}
}

\section{Conclusion}
We study how \CFA evidence can be used to identify corrupted branch targets, pinpoint exploited memory instructions, and generate/verify binary patches. We analyze and classify types of \CFA evidence and conclude that verbatim evidence provides \vrf with the most flexibility for remote analysis. We introduce \acron to verify path validity and to locate/patch vulnerabilities using only the application binary and \CFA evidence. We implement and evaluate \acron's public  prototype~\cite{repo}, which demonstrates its effectiveness atop hardware-based and TEE-based \CFA architectures.

\vspace{2mm}
\noindent{\bf Acknowledgments.} We thank the anonymous reviewers and shepherd for their guidance.

\bibliographystyle{ACM-Reference-Format}
% \bibliography{references}
\bibliography{small_refs}


\begin{thebibliography}{89}

%%% the Web address.

\ifx \showCODEN    \undefined \def \showCODEN     #1{\unskip}     \fi
\ifx \showDOI      \undefined \def \showDOI       #1{#1}\fi
\ifx \showISBNx    \undefined \def \showISBNx     #1{\unskip}     \fi
\ifx \showISBNxiii \undefined \def \showISBNxiii  #1{\unskip}     \fi
\ifx \showISSN     \undefined \def \showISSN      #1{\unskip}     \fi
\ifx \showLCCN     \undefined \def \showLCCN      #1{\unskip}     \fi
\ifx \shownote     \undefined \def \shownote      #1{#1}          \fi
\ifx \showarticletitle \undefined \def \showarticletitle #1{#1}   \fi
\ifx \showURL      \undefined \def \showURL       {\relax}        \fi
% invisible to TeX
\providecommand\bibfield[2]{#2}
\providecommand\bibinfo[2]{#2}
\providecommand\natexlab[1]{#1}
\providecommand\showeprint[2][]{arXiv:#2}

\bibitem[Abera et~al\mbox{.}(2016)]%
        {cflat}
\bibfield{author}{\bibinfo{person}{Tigist Abera} {et~al\mbox{.}}}
  \bibinfo{year}{2016}\natexlab{}.
\newblock \showarticletitle{{C-FLAT}: control-flow attestation for embedded
  systems software}. In \bibinfo{booktitle}{\emph{Proceedings of the 2016 ACM
  SIGSAC Conference on Computer and Communications Security (CCS)}}.
  \bibinfo{pages}{743--754}.
\newblock


\bibitem[Ammar et~al\mbox{.}(2024)]%
        {sok_cfa_cfi}
\bibfield{author}{\bibinfo{person}{Mahmoud Ammar} {et~al\mbox{.}}}
  \bibinfo{year}{2024}\natexlab{}.
\newblock \showarticletitle{SoK: Integrity, Attestation, and Auditing of
  Program Execution}. In \bibinfo{booktitle}{\emph{2025 IEEE Symposium on
  Security and Privacy (SP)}}. IEEE Computer Society, \bibinfo{pages}{77--77}.
\newblock


\bibitem[Armanuzzaman et~al\mbox{.}(2025)]%
        {enola}
\bibfield{author}{\bibinfo{person}{Md Armanuzzaman} {et~al\mbox{.}}}
  \bibinfo{year}{2025}\natexlab{}.
\newblock \showarticletitle{{ENOLA}: Efficient Control-Flow Attestation for
  Embedded Systems}.
\newblock \bibinfo{journal}{\emph{arXiv preprint arXiv:2501.11207}}
  (\bibinfo{year}{2025}).
\newblock


\bibitem[Baldoni et~al\mbox{.}(2018)]%
        {baldoni2018survey}
\bibfield{author}{\bibinfo{person}{Roberto Baldoni} {et~al\mbox{.}}}
  \bibinfo{year}{2018}\natexlab{}.
\newblock \showarticletitle{A survey of symbolic execution techniques}.
\newblock \bibinfo{journal}{\emph{ACM Computing Surveys (CSUR)}}
  \bibinfo{volume}{51}, \bibinfo{number}{3} (\bibinfo{year}{2018}),
  \bibinfo{pages}{1--39}.
\newblock


\bibitem[Bletsch et~al\mbox{.}(2011)]%
        {jop}
\bibfield{author}{\bibinfo{person}{Tyler Bletsch} {et~al\mbox{.}}}
  \bibinfo{year}{2011}\natexlab{}.
\newblock \showarticletitle{Jump-oriented programming: a new class of
  code-reuse attack}. In \bibinfo{booktitle}{\emph{Proceedings of the 6th ACM
  symposium on information, computer and communications security (CCS)}}.
  \bibinfo{pages}{30--40}.
\newblock


\bibitem[Brasser et~al\mbox{.}(2015)]%
        {tytan}
\bibfield{author}{\bibinfo{person}{Ferdinand Brasser} {et~al\mbox{.}}}
  \bibinfo{year}{2015}\natexlab{}.
\newblock \showarticletitle{{TyTAN}: Tiny trust anchor for tiny devices}. In
  \bibinfo{booktitle}{\emph{Proceedings of the 52nd annual design automation
  conference (DAC)}}. \bibinfo{pages}{1--6}.
\newblock


\bibitem[Burow et~al\mbox{.}(2019)]%
        {burow2019sok}
\bibfield{author}{\bibinfo{person}{Nathan Burow} {et~al\mbox{.}}}
  \bibinfo{year}{2019}\natexlab{}.
\newblock \showarticletitle{SoK: Shining light on shadow stacks}. In
  \bibinfo{booktitle}{\emph{IEEE Symposium on Security and Privacy (SP)}}.
  IEEE, \bibinfo{pages}{985--999}.
\newblock


\bibitem[Caulfield et~al\mbox{.}(2022)]%
        {asap}
\bibfield{author}{\bibinfo{person}{Adam Caulfield} {et~al\mbox{.}}}
  \bibinfo{year}{2022}\natexlab{}.
\newblock \showarticletitle{{ASAP}: reconciling asynchronous real-time
  operations and proofs of execution in simple embedded systems}. In
  \bibinfo{booktitle}{\emph{Proceedings of the 59th ACM/IEEE Design Automation
  Conference (DAC)}}. \bibinfo{pages}{721--726}.
\newblock


\bibitem[Caulfield et~al\mbox{.}(2024a)]%
        {speccfa}
\bibfield{author}{\bibinfo{person}{Adam Caulfield} {et~al\mbox{.}}}
  \bibinfo{year}{2024}\natexlab{a}.
\newblock \showarticletitle{{SpecCFA:} Enhancing Control Flow
  Attestation/Auditing via Application-Aware Sub-Path Speculation}.
\newblock  (\bibinfo{year}{2024}), \bibinfo{pages}{563--578}.
\newblock


\bibitem[Caulfield et~al\mbox{.}(2024b)]%
        {traces}
\bibfield{author}{\bibinfo{person}{Adam Caulfield} {et~al\mbox{.}}}
  \bibinfo{year}{2024}\natexlab{b}.
\newblock \showarticletitle{{TRACES:} TEE-based Runtime Auditing for Commodity
  Embedded Systems}.
\newblock  (\bibinfo{year}{2024}), \bibinfo{pages}{257--270}.
\newblock


\bibitem[Caulfield et~al\mbox{.}(2025)]%
        {repo}
\bibfield{author}{\bibinfo{person}{Adam Caulfield} {et~al\mbox{.}}}
  \bibinfo{year}{2025}\natexlab{}.
\newblock \bibinfo{title}{{Github Repository for SABRE Prototype}}.
\newblock \bibinfo{howpublished}{\url{https://github.com/SPINS-RG/SABRE}}.
\newblock


\bibitem[Caulfield and otehrs(2023)]%
        {acfa}
\bibfield{author}{\bibinfo{person}{Adam Caulfield} {and}
  \bibinfo{person}{otehrs}.} \bibinfo{year}{2023}\natexlab{}.
\newblock \showarticletitle{{ACFA}: Secure Runtime Auditing \& Guaranteed
  Device Healing via Active Control Flow Attestation}. In
  \bibinfo{booktitle}{\emph{32nd USENIX Security Symposium}}.
  \bibinfo{publisher}{USENIX Association}, \bibinfo{pages}{5827--5844}.
\newblock


\bibitem[Cheng et~al\mbox{.}(2019)]%
        {data_oriented_attacks}
\bibfield{author}{\bibinfo{person}{Long Cheng} {et~al\mbox{.}}}
  \bibinfo{year}{2019}\natexlab{}.
\newblock \showarticletitle{Exploitation techniques and defenses for
  data-oriented attacks}. In \bibinfo{booktitle}{\emph{2019 IEEE Cybersecurity
  Development (SecDev)}}. IEEE, \bibinfo{pages}{114--128}.
\newblock


\bibitem[Chilese et~al\mbox{.}(2024)]%
        {rage}
\bibfield{author}{\bibinfo{person}{Marco Chilese} {et~al\mbox{.}}}
  \bibinfo{year}{2024}\natexlab{}.
\newblock \showarticletitle{One for All and All for One: GNN-based Control-Flow
  Attestation for Embedded Devices}. In \bibinfo{booktitle}{\emph{IEEE
  Symposium on Security and Privacy (SP)}}. IEEE, \bibinfo{pages}{203--203}.
\newblock


\bibitem[(CISA)(2023)]%
        {cisa-urgent-need-for-memory-safety}
\bibfield{author}{\bibinfo{person}{Bob~Lord (CISA)}.}
  \bibinfo{year}{2023}\natexlab{}.
\newblock \bibinfo{title}{The Urgent Need for Memory Safety in Software
  Products}.
\newblock
  \bibinfo{howpublished}{\url{https://www.cisa.gov/news-events/news/urgent-need-memory-safety-software-products}}.
\newblock


\bibitem[Coker et~al\mbox{.}(2011)]%
        {coker2011principles}
\bibfield{author}{\bibinfo{person}{George Coker} {et~al\mbox{.}}}
  \bibinfo{year}{2011}\natexlab{}.
\newblock \showarticletitle{Principles of remote attestation}.
\newblock \bibinfo{journal}{\emph{International Journal of Information
  Security}}  \bibinfo{volume}{10} (\bibinfo{year}{2011}),
  \bibinfo{pages}{63--81}.
\newblock


\bibitem[Cowan et~al\mbox{.}(2000)]%
        {cowan2000buffer}
\bibfield{author}{\bibinfo{person}{Crispin Cowan} {et~al\mbox{.}}}
  \bibinfo{year}{2000}\natexlab{}.
\newblock \showarticletitle{Buffer overflows: Attacks and defenses for the
  vulnerability of the decade}. In \bibinfo{booktitle}{\emph{Proceedings DARPA
  Information Survivability Conference and Exposition (DISCEX)}},
  Vol.~\bibinfo{volume}{2}. IEEE, \bibinfo{pages}{119--129}.
\newblock


\bibitem[Debes et~al\mbox{.}(2023)]%
        {debes2023zekra}
\bibfield{author}{\bibinfo{person}{Heini~Bergsson Debes} {et~al\mbox{.}}}
  \bibinfo{year}{2023}\natexlab{}.
\newblock \showarticletitle{{ZEKRA}: Zero-Knowledge Control-Flow Attestation}.
  In \bibinfo{booktitle}{\emph{Proceedings of the 2023 ACM Asia Conference on
  Computer and Communications Security (AsiaCCS)}}. \bibinfo{pages}{357--371}.
\newblock


\bibitem[Deligiannis and Kornaros(2016)]%
        {deligiannis2016adaptive}
\bibfield{author}{\bibinfo{person}{Ioannis Deligiannis} {and}
  \bibinfo{person}{George Kornaros}.} \bibinfo{year}{2016}\natexlab{}.
\newblock \showarticletitle{Adaptive memory management scheme for MMU-less
  embedded systems}. In \bibinfo{booktitle}{\emph{11th Symposium on Industrial
  Embedded Systems (SIES)}}. IEEE, \bibinfo{pages}{1--8}.
\newblock


\bibitem[Dessouky et~al\mbox{.}(2017)]%
        {lofat}
\bibfield{author}{\bibinfo{person}{Ghada Dessouky} {et~al\mbox{.}}}
  \bibinfo{year}{2017}\natexlab{}.
\newblock \showarticletitle{{LO-FAT}: Low-overhead control flow attestation in
  hardware}. In \bibinfo{booktitle}{\emph{Proceedings of the 54th Annual Design
  Automation Conference (DAC)}}. \bibinfo{pages}{1--6}.
\newblock


\bibitem[Dessouky et~al\mbox{.}(2018)]%
        {litehax}
\bibfield{author}{\bibinfo{person}{Ghada Dessouky} {et~al\mbox{.}}}
  \bibinfo{year}{2018}\natexlab{}.
\newblock \showarticletitle{{Litehax}: lightweight hardware-assisted
  attestation of program execution}. In \bibinfo{booktitle}{\emph{International
  Conference on Computer-Aided Design (ICCAD)}}. IEEE, \bibinfo{pages}{1--8}.
\newblock


\bibitem[Duan et~al\mbox{.}(2019)]%
        {osspatcher}
\bibfield{author}{\bibinfo{person}{Ruian Duan} {et~al\mbox{.}}}
  \bibinfo{year}{2019}\natexlab{}.
\newblock \showarticletitle{Automating Patching of Vulnerable Open-Source
  Software Versions in Application Binaries.}. In
  \bibinfo{booktitle}{\emph{Network and Distributed System Security (NDSS)
  Symposium}}.
\newblock


\bibitem[Duck et~al\mbox{.}(2020)]%
        {e9patch}
\bibfield{author}{\bibinfo{person}{Gregory~J Duck} {et~al\mbox{.}}}
  \bibinfo{year}{2020}\natexlab{}.
\newblock \showarticletitle{Binary rewriting without control flow recovery}. In
  \bibinfo{booktitle}{\emph{Proceedings of the 41st ACM SIGPLAN conference on
  programming language design and implementation (PLDI)}}.
  \bibinfo{pages}{151--163}.
\newblock


\bibitem[Eldefrawy et~al\mbox{.}(2012)]%
        {smart}
\bibfield{author}{\bibinfo{person}{Karim Eldefrawy} {et~al\mbox{.}}}
  \bibinfo{year}{2012}\natexlab{}.
\newblock \showarticletitle{{SMART}: Secure and Minimal Architecture for
  (Establishing Dynamic) Root of Trust}. In \bibinfo{booktitle}{\emph{Network
  and Distributed System Security (NDSS) Symposium}},
  Vol.~\bibinfo{volume}{12}. \bibinfo{pages}{1--15}.
\newblock


\bibitem[Eldefrawy et~al\mbox{.}(2020)]%
        {alice}
\bibfield{author}{\bibinfo{person}{Karim Eldefrawy}, \bibinfo{person}{Michael
  Locasto}, \bibinfo{person}{Norrathep Rattanavipanon}, {and}
  \bibinfo{person}{Hassen Saidi}.} \bibinfo{year}{2020}\natexlab{}.
\newblock \showarticletitle{Towards Automated Augmentation and Instrumentation
  of Legacy Cryptographic Executables}. In \bibinfo{booktitle}{\emph{Applied
  Cryptography and Network Security (ACNS)}}. Springer,
  \bibinfo{pages}{364--384}.
\newblock


\bibitem[Foundation(2024a)]%
        {arm_objdump}
\bibfield{author}{\bibinfo{person}{Free~Software Foundation}.}
  \bibinfo{year}{2024}\natexlab{a}.
\newblock \bibinfo{title}{{arm-none-eabi-objdump man page}}.
\newblock
\newblock
\urldef\tempurl%
\url{https://manpages.debian.org/unstable/binutils-arm-none-eabi/arm-none-eabi-objdump.1.en.html}
\showURL{%
\tempurl}


\bibitem[Foundation(2024b)]%
        {msp430_objdump}
\bibfield{author}{\bibinfo{person}{Free~Software Foundation}.}
  \bibinfo{year}{2024}\natexlab{b}.
\newblock \bibinfo{title}{{msp430-objdump man page}}.
\newblock
\newblock
\urldef\tempurl%
\url{https://manpages.debian.org/testing/binutils-msp430/msp430-objdump.1.en.html}
\showURL{%
\tempurl}


\bibitem[Foundation(2024c)]%
        {pyelftools}
\bibfield{author}{\bibinfo{person}{Python~Software Foundation}.}
  \bibinfo{year}{2024}\natexlab{c}.
\newblock \bibinfo{title}{{Python libary pyelftools}}.
\newblock
\newblock
\urldef\tempurl%
\url{https://pypi.org/project/pyelftools/0.20/}
\showURL{%
\tempurl}


\bibitem[Geden and Rasmussen(2019)]%
        {geden2019hardware}
\bibfield{author}{\bibinfo{person}{Munir Geden} {and} \bibinfo{person}{Kasper
  Rasmussen}.} \bibinfo{year}{2019}\natexlab{}.
\newblock \showarticletitle{Hardware-assisted remote runtime attestation for
  critical embedded systems}. In \bibinfo{booktitle}{\emph{17th International
  Conference on Privacy, Security and Trust (PST)}}. IEEE,
  \bibinfo{pages}{1--10}.
\newblock


\bibitem[{Intel}(2015)]%
        {intel_pt}
\bibfield{author}{\bibinfo{person}{{Intel}}.} \bibinfo{year}{2015}\natexlab{}.
\newblock \bibinfo{title}{{Intel Processor Trace}}.
\newblock
  \bibinfo{howpublished}{\url{https://edc.intel.com/content/www/us/en/design/ipla/software-development-platforms/client/platforms/alder-lake-desktop/12th-generation-intel-core-processors-datasheet-volume-1-of-2/010/intel-processor-trace/}}.
\newblock
\newblock
\shownote{[Online; accessed 14-March-2025]}.


\bibitem[Kayan et~al\mbox{.}(2022)]%
        {kayan2022cybersecurity}
\bibfield{author}{\bibinfo{person}{Hakan Kayan} {et~al\mbox{.}}}
  \bibinfo{year}{2022}\natexlab{}.
\newblock \showarticletitle{Cybersecurity of industrial cyber-physical systems:
  a review}.
\newblock \bibinfo{journal}{\emph{ACM Computing Surveys (CSUR)}}
  \bibinfo{volume}{54}, \bibinfo{number}{11s} (\bibinfo{year}{2022}),
  \bibinfo{pages}{1--35}.
\newblock


\bibitem[Kennell and Jamieson(2003)]%
        {kennell2003establish}
\bibfield{author}{\bibinfo{person}{Rick Kennell} {and} \bibinfo{person}{Leah~H
  Jamieson}.} \bibinfo{year}{2003}\natexlab{}.
\newblock \showarticletitle{Establishing the genuinity of remote computer
  systems}. In \bibinfo{booktitle}{\emph{12th USENIX Security Symposium}}.
\newblock


\bibitem[Keystone(2024)]%
        {keystone-engine}
\bibfield{author}{\bibinfo{person}{Keystone}.} \bibinfo{year}{2024}\natexlab{}.
\newblock \bibinfo{title}{{Keystone: the ultimate assembler}}.
\newblock
\newblock
\urldef\tempurl%
\url{https://www.keystone-engine.org/}
\showURL{%
\tempurl}


\bibitem[Kiaei et~al\mbox{.}(2021)]%
        {fi_safe_bin}
\bibfield{author}{\bibinfo{person}{Pantea Kiaei} {et~al\mbox{.}}}
  \bibinfo{year}{2021}\natexlab{}.
\newblock \showarticletitle{Rewrite to reinforce: Rewriting the binary to apply
  countermeasures against fault injection}. In \bibinfo{booktitle}{\emph{58th
  Design Automation Conference (DAC)}}. IEEE, \bibinfo{pages}{319--324}.
\newblock


\bibitem[King(1976)]%
        {king1976symbolic}
\bibfield{author}{\bibinfo{person}{James~C King}.}
  \bibinfo{year}{1976}\natexlab{}.
\newblock \showarticletitle{Symbolic execution and program testing}.
\newblock \bibinfo{journal}{\emph{Commun. ACM}} \bibinfo{volume}{19},
  \bibinfo{number}{7} (\bibinfo{year}{1976}), \bibinfo{pages}{385--394}.
\newblock


\bibitem[Kovah et~al\mbox{.}(2012)]%
        {checkmate_att}
\bibfield{author}{\bibinfo{person}{Xeno Kovah} {et~al\mbox{.}}}
  \bibinfo{year}{2012}\natexlab{}.
\newblock \showarticletitle{New results for timing-based attestation}. In
  \bibinfo{booktitle}{\emph{IEEE Symposium on Security and Privacy (SP)}}.
  IEEE, \bibinfo{pages}{239--253}.
\newblock


\bibitem[Liu et~al\mbox{.}(2022)]%
        {liu2022seeker}
\bibfield{author}{\bibinfo{person}{Runhao Liu} {et~al\mbox{.}}}
  \bibinfo{year}{2022}\natexlab{}.
\newblock \showarticletitle{{SEEKER}: A root cause analysis method based on
  deterministic replay for multi-type network protocol vulnerabilities}. In
  \bibinfo{booktitle}{\emph{International Conference on Trust, Security and
  Privacy in Computing and Communications (TrustCom)}}. IEEE,
  \bibinfo{pages}{131--138}.
\newblock


\bibitem[Masmano et~al\mbox{.}(2003)]%
        {masmano2003dynamic}
\bibfield{author}{\bibinfo{person}{Miguel Masmano} {et~al\mbox{.}}}
  \bibinfo{year}{2003}\natexlab{}.
\newblock \showarticletitle{Dynamic storage allocation for real-time embedded
  systems}.
\newblock \bibinfo{journal}{\emph{Proc. of Real-Time System Simposium WIP}}
  (\bibinfo{year}{2003}).
\newblock


\bibitem[(MITRE)(2024)]%
        {mitre-cwe-top-25-2024}
\bibfield{author}{\bibinfo{person}{The MITRE~Corporation (MITRE)}.}
  \bibinfo{year}{2024}\natexlab{}.
\newblock \bibinfo{title}{2024 CWE Top 25 Most Dangerous Software Weaknesses}.
\newblock
  \bibinfo{howpublished}{\url{https://cwe.mitre.org/top25/archive/2024/2024_cwe_top25.html}}.
\newblock


\bibitem[Nafees et~al\mbox{.}(2023)]%
        {nafees2023smart}
\bibfield{author}{\bibinfo{person}{Muhammad~Nouman Nafees} {et~al\mbox{.}}}
  \bibinfo{year}{2023}\natexlab{}.
\newblock \showarticletitle{Smart grid cyber-physical situational awareness of
  complex operational technology attacks: A review}.
\newblock \bibinfo{journal}{\emph{Comput. Surveys}} \bibinfo{volume}{55},
  \bibinfo{number}{10} (\bibinfo{year}{2023}), \bibinfo{pages}{1--36}.
\newblock


\bibitem[{National Vulnerability Database}(2017)]%
        {cve_2017_14201}
\bibfield{author}{\bibinfo{person}{{National Vulnerability Database}}.}
  \bibinfo{year}{2017}\natexlab{}.
\newblock \bibinfo{title}{{CVE-2017-14201}}.
\newblock
\newblock
\urldef\tempurl%
\url{https://nvd.nist.gov/vuln/detail/CVE-2017-14201}
\showURL{%
\tempurl}
\newblock
\shownote{Accessed: 2025-May-04}.


\bibitem[{National Vulnerability Database}(2020a)]%
        {cve_2019_16127}
\bibfield{author}{\bibinfo{person}{{National Vulnerability Database}}.}
  \bibinfo{year}{2020}\natexlab{a}.
\newblock \bibinfo{title}{{CVE-2019-16127}}.
\newblock
\newblock
\urldef\tempurl%
\url{https://nvd.nist.gov/vuln/detail/CVE-2019-16127}
\showURL{%
\tempurl}
\newblock
\shownote{Accessed: 2025-May-04}.


\bibitem[{National Vulnerability Database}(2020b)]%
        {cve_2020_10019}
\bibfield{author}{\bibinfo{person}{{National Vulnerability Database}}.}
  \bibinfo{year}{2020}\natexlab{b}.
\newblock \bibinfo{title}{{CVE-2020-10019}}.
\newblock
\newblock
\urldef\tempurl%
\url{https://nvd.nist.gov/vuln/detail/CVE-2020-10019}
\showURL{%
\tempurl}
\newblock
\shownote{Accessed: 2025-May-04}.


\bibitem[{National Vulnerability Database}(2020c)]%
        {cve_2020_10023}
\bibfield{author}{\bibinfo{person}{{National Vulnerability Database}}.}
  \bibinfo{year}{2020}\natexlab{c}.
\newblock \bibinfo{title}{{CVE-2020-10023}}.
\newblock
\newblock
\urldef\tempurl%
\url{https://nvd.nist.gov/vuln/detail/CVE-2020-10023}
\showURL{%
\tempurl}
\newblock
\shownote{Accessed: 2025-May-04}.


\bibitem[{National Vulnerability Database}(2021a)]%
        {cve_2021_0920}
\bibfield{author}{\bibinfo{person}{{National Vulnerability Database}}.}
  \bibinfo{year}{2021}\natexlab{a}.
\newblock \bibinfo{title}{{CVE-2021-0920}}.
\newblock
\newblock
\urldef\tempurl%
\url{https://nvd.nist.gov/vuln/detail/CVE-2021-0920}
\showURL{%
\tempurl}
\newblock
\shownote{Accessed: 2025-May-04}.


\bibitem[{National Vulnerability Database}(2021b)]%
        {cve_2021_35395}
\bibfield{author}{\bibinfo{person}{{National Vulnerability Database}}.}
  \bibinfo{year}{2021}\natexlab{b}.
\newblock \bibinfo{title}{{CVE-2021-35395}}.
\newblock
\newblock
\urldef\tempurl%
\url{https://nvd.nist.gov/vuln/detail/CVE-2021-35395}
\showURL{%
\tempurl}
\newblock
\shownote{Accessed: 2025-May-04}.


\bibitem[{National Vulnerability Database}(2022)]%
        {cve_2022_34835}
\bibfield{author}{\bibinfo{person}{{National Vulnerability Database}}.}
  \bibinfo{year}{2022}\natexlab{}.
\newblock \bibinfo{title}{{CVE-2022-34835}}.
\newblock
\newblock
\urldef\tempurl%
\url{https://nvd.nist.gov/vuln/detail/CVE-2022-34835}
\showURL{%
\tempurl}
\newblock
\shownote{Accessed: 2025-May-04}.


\bibitem[Neto and Nunes(2023)]%
        {iscflat}
\bibfield{author}{\bibinfo{person}{Antonio~Joia Neto} {and}
  \bibinfo{person}{Ivan De~Oliveira Nunes}.} \bibinfo{year}{2023}\natexlab{}.
\newblock \showarticletitle{{ISC-FLAT}: On the Conflict Between Control Flow
  Attestation and Real-Time Operations}. In \bibinfo{booktitle}{\emph{29th
  Real-Time and Embedded Technology and Applications Symposium (RTAS)}}. IEEE,
  \bibinfo{pages}{133--146}.
\newblock


\bibitem[Noorman et~al\mbox{.}(2017)]%
        {Sancus17}
\bibfield{author}{\bibinfo{person}{Job Noorman} {et~al\mbox{.}}}
  \bibinfo{year}{2017}\natexlab{}.
\newblock \showarticletitle{Sancus 2.0: A low-cost security architecture for
  iot devices}.
\newblock \bibinfo{journal}{\emph{ACM Transactions on Privacy and Security
  (TOPS)}} \bibinfo{volume}{20}, \bibinfo{number}{3} (\bibinfo{year}{2017}),
  \bibinfo{pages}{1--33}.
\newblock


\bibitem[Nunes et~al\mbox{.}(2019)]%
        {vrased}
\bibfield{author}{\bibinfo{person}{Ivan De~Oliveira Nunes} {et~al\mbox{.}}}
  \bibinfo{year}{2019}\natexlab{}.
\newblock \showarticletitle{{VRASED}: A Verified {Hardware/Software Co-Design}
  for Remote Attestation}. In \bibinfo{booktitle}{\emph{28th USENIX Security
  Symposium}}. \bibinfo{pages}{1429--1446}.
\newblock


\bibitem[Nunes et~al\mbox{.}(2020)]%
        {apex}
\bibfield{author}{\bibinfo{person}{Ivan De~Oliveira Nunes} {et~al\mbox{.}}}
  \bibinfo{year}{2020}\natexlab{}.
\newblock \showarticletitle{{APEX}: A verified architecture for proofs of
  execution on remote devices under full software compromise}. In
  \bibinfo{booktitle}{\emph{29th USENIX Security Symposium}}.
  \bibinfo{pages}{771--788}.
\newblock


\bibitem[Nunes et~al\mbox{.}(2021a)]%
        {dialed}
\bibfield{author}{\bibinfo{person}{Ivan De~Oliveira Nunes} {et~al\mbox{.}}}
  \bibinfo{year}{2021}\natexlab{a}.
\newblock \showarticletitle{{Dialed}: Data integrity attestation for low-end
  embedded devices}. In \bibinfo{booktitle}{\emph{58th Design Automation
  Conference (DAC)}}. IEEE, \bibinfo{pages}{313--318}.
\newblock


\bibitem[Nunes et~al\mbox{.}(2021b)]%
        {tinycfa}
\bibfield{author}{\bibinfo{person}{Ivan De~Oliveira Nunes} {et~al\mbox{.}}}
  \bibinfo{year}{2021}\natexlab{b}.
\newblock \showarticletitle{{Tiny-CFA}: Minimalistic control-flow attestation
  using verified proofs of execution}. In \bibinfo{booktitle}{\emph{2021
  Design, Automation \& Test in Europe Conference \& Exhibition (DATE)}}. IEEE,
  \bibinfo{pages}{641--646}.
\newblock


\bibitem[Pallister et~al\mbox{.}(2013)]%
        {beebs}
\bibfield{author}{\bibinfo{person}{James Pallister} {et~al\mbox{.}}}
  \bibinfo{year}{2013}\natexlab{}.
\newblock \showarticletitle{BEEBS: Open benchmarks for energy measurements on
  embedded platforms}.
\newblock \bibinfo{journal}{\emph{arXiv preprint arXiv:1308.5174}}
  (\bibinfo{year}{2013}).
\newblock


\bibitem[Park et~al\mbox{.}(2024)]%
        {park2024benzene}
\bibfield{author}{\bibinfo{person}{Younggi Park} {et~al\mbox{.}}}
  \bibinfo{year}{2024}\natexlab{}.
\newblock \showarticletitle{BENZENE: A Practical Root Cause Analysis System
  with an Under-Constrained State Mutation}. In \bibinfo{booktitle}{\emph{IEEE
  Symposium on Security and Privacy (SP)}}. IEEE, \bibinfo{pages}{1865--1883}.
\newblock


\bibitem[Petroni~Jr et~al\mbox{.}(2004)]%
        {copilot}
\bibfield{author}{\bibinfo{person}{Nick~L Petroni~Jr} {et~al\mbox{.}}}
  \bibinfo{year}{2004}\natexlab{}.
\newblock \showarticletitle{Copilot-a coprocessor-based kernel runtime
  integrity monitor.}. In \bibinfo{booktitle}{\emph{USENIX security
  symposium}}. San Diego, USA, \bibinfo{pages}{179--194}.
\newblock


\bibitem[Rajput et~al\mbox{.}(2023)]%
        {icspatch}
\bibfield{author}{\bibinfo{person}{Prashant Hari~Narayan Rajput}
  {et~al\mbox{.}}} \bibinfo{year}{2023}\natexlab{}.
\newblock \showarticletitle{{ICSPatch}: Automated Vulnerability Localization
  and $\{$Non-Intrusive$\}$ Hotpatching in Industrial Control Systems using
  Data Dependence Graphs}. In \bibinfo{booktitle}{\emph{32nd USENIX Security
  Symposium}}. \bibinfo{pages}{6861--6876}.
\newblock


\bibitem[Ramakrishna et~al\mbox{.}(2008)]%
        {ramakrishna2008smart}
\bibfield{author}{\bibinfo{person}{M Ramakrishna} {et~al\mbox{.}}}
  \bibinfo{year}{2008}\natexlab{}.
\newblock \showarticletitle{Smart dynamic memory allocator for embedded
  systems}. In \bibinfo{booktitle}{\emph{23rd International Symposium on
  Computer and Information Sciences (ISCIS)}}. IEEE, \bibinfo{pages}{1--6}.
\newblock


\bibitem[Ramalingam(1994)]%
        {aliasing}
\bibfield{author}{\bibinfo{person}{Ganesan Ramalingam}.}
  \bibinfo{year}{1994}\natexlab{}.
\newblock \showarticletitle{The undecidability of aliasing}.
\newblock \bibinfo{journal}{\emph{ACM Transactions on Programming Languages and
  Systems (TOPLAS)}} \bibinfo{volume}{16}, \bibinfo{number}{5}
  (\bibinfo{year}{1994}), \bibinfo{pages}{1467--1471}.
\newblock


\bibitem[{Realtek Semiconductor Corp.}(2021)]%
        {realtek_vuln_report}
\bibfield{author}{\bibinfo{person}{{Realtek Semiconductor Corp.}}}
  \bibinfo{year}{2021}\natexlab{}.
\newblock \bibinfo{title}{{Realtek AP-Router SDK Advisory}}.
\newblock
\newblock
\urldef\tempurl%
\url{https://www.realtek.com/images/safe-report/Realtek_APRouter_SDK_Advisory-CVE-2021-35392_35395.pdf}
\showURL{%
\tempurl}
\newblock
\shownote{Accessed: 2025-May-04}.


\bibitem[Roemer et~al\mbox{.}(2012)]%
        {rop}
\bibfield{author}{\bibinfo{person}{Ryan Roemer} {et~al\mbox{.}}}
  \bibinfo{year}{2012}\natexlab{}.
\newblock \showarticletitle{Return-oriented programming: Systems, languages,
  and applications}.
\newblock \bibinfo{journal}{\emph{ACM Transactions on Information and System
  Security (TISSEC)}} \bibinfo{volume}{15}, \bibinfo{number}{1}
  (\bibinfo{year}{2012}), \bibinfo{pages}{1--34}.
\newblock


\bibitem[Schellekens et~al\mbox{.}(2008)]%
        {tpm_attest}
\bibfield{author}{\bibinfo{person}{Dries Schellekens}, \bibinfo{person}{Brecht
  Wyseur}, {and} \bibinfo{person}{Bart Preneel}.}
  \bibinfo{year}{2008}\natexlab{}.
\newblock \showarticletitle{Remote attestation on legacy operating systems with
  trusted platform modules}.
\newblock \bibinfo{journal}{\emph{Science of Computer Programming}}
  \bibinfo{volume}{74}, \bibinfo{number}{1-2} (\bibinfo{year}{2008}),
  \bibinfo{pages}{13--22}.
\newblock


\bibitem[Seshadri et~al\mbox{.}(2004)]%
        {swatt}
\bibfield{author}{\bibinfo{person}{Arvind Seshadri} {et~al\mbox{.}}}
  \bibinfo{year}{2004}\natexlab{}.
\newblock \showarticletitle{{SWATT}: Software-based attestation for embedded
  devices}. In \bibinfo{booktitle}{\emph{IEEE Symposium on Security and Privacy
  (SP)}}. IEEE, \bibinfo{pages}{272--282}.
\newblock


\bibitem[Seshadri et~al\mbox{.}(2005)]%
        {pioneer}
\bibfield{author}{\bibinfo{person}{Arvind Seshadri} {et~al\mbox{.}}}
  \bibinfo{year}{2005}\natexlab{}.
\newblock \showarticletitle{{Pioneer}: verifying code integrity and enforcing
  untampered code execution on legacy systems}. In
  \bibinfo{booktitle}{\emph{Proceedings of the twentieth ACM symposium on
  Operating Systems Principles (SOSP)}}. \bibinfo{pages}{1--16}.
\newblock


\bibitem[Seshadri et~al\mbox{.}(2008)]%
        {sake}
\bibfield{author}{\bibinfo{person}{Arvind Seshadri} {et~al\mbox{.}}}
  \bibinfo{year}{2008}\natexlab{}.
\newblock \showarticletitle{{SAKE}: Software attestation for key establishment
  in sensor networks}.
\newblock In \bibinfo{booktitle}{\emph{Distributed Computing in Sensor Systems
  (DCOSS)}}. \bibinfo{pages}{372--385}.
\newblock


\bibitem[Shacham et~al\mbox{.}(2004)]%
        {aslr}
\bibfield{author}{\bibinfo{person}{Hovav Shacham} {et~al\mbox{.}}}
  \bibinfo{year}{2004}\natexlab{}.
\newblock \showarticletitle{On the effectiveness of address-space
  randomization}. In \bibinfo{booktitle}{\emph{Proceedings of the 11th ACM
  conference on Computer and communications security (CCS)}}.
  \bibinfo{pages}{298--307}.
\newblock


\bibitem[Sun et~al\mbox{.}(2020)]%
        {oat}
\bibfield{author}{\bibinfo{person}{Zhichuang Sun} {et~al\mbox{.}}}
  \bibinfo{year}{2020}\natexlab{}.
\newblock \showarticletitle{{OAT}: Attesting operation integrity of embedded
  devices}. In \bibinfo{booktitle}{\emph{IEEE Symposium on Security and Privacy
  (SP)}}. IEEE, \bibinfo{pages}{1433--1449}.
\newblock


\bibitem[Swiftloke(2024)]%
        {msprobe}
\bibfield{author}{\bibinfo{person}{Swiftloke}.}
  \bibinfo{year}{2024}\natexlab{}.
\newblock \bibinfo{title}{{Github Repository for MSProbe}}.
\newblock
\newblock
\urldef\tempurl%
\url{https://github.com/Swiftloke/MSProbe}
\showURL{%
\tempurl}


\bibitem[Tan et~al\mbox{.}(2024)]%
        {tan2024sok}
\bibfield{author}{\bibinfo{person}{Xi Tan} {et~al\mbox{.}}}
  \bibinfo{year}{2024}\natexlab{}.
\newblock \showarticletitle{{SoK}:{Where’s} the {“up”?!} A Comprehensive
  (bottom-up) Study on the Security of Arm Cortex-M Systems}. In
  \bibinfo{booktitle}{\emph{18th USENIX WOOT Conference on Offensive
  Technologies (WOOT)}}. \bibinfo{pages}{149--169}.
\newblock


\bibitem[Team(2024)]%
        {sympy}
\bibfield{author}{\bibinfo{person}{SymPy~Development Team}.}
  \bibinfo{year}{2024}\natexlab{}.
\newblock \bibinfo{title}{{sympy}}.
\newblock
\newblock
\urldef\tempurl%
\url{https://docs.sympy.org/latest/index.html}
\showURL{%
\tempurl}


\bibitem[Toffalini et~al\mbox{.}(2019)]%
        {scarr}
\bibfield{author}{\bibinfo{person}{Flavio Toffalini} {et~al\mbox{.}}}
  \bibinfo{year}{2019}\natexlab{}.
\newblock \showarticletitle{{ScaRR}: Scalable Runtime Remote Attestation for
  Complex Systems}. In \bibinfo{booktitle}{\emph{22nd International Symposium
  on Research in Attacks, Intrusions and Defenses (RAID)}}.
  \bibinfo{pages}{121--134}.
\newblock


\bibitem[Vliegen et~al\mbox{.}(2019)]%
        {sacha}
\bibfield{author}{\bibinfo{person}{Jo Vliegen} {et~al\mbox{.}}}
  \bibinfo{year}{2019}\natexlab{}.
\newblock \showarticletitle{{SACHa}: Self-attestation of configurable
  hardware}. In \bibinfo{booktitle}{\emph{Design, Automation \& Test in Europe
  Conference \& Exhibition (DATE)}}. IEEE, \bibinfo{pages}{746--751}.
\newblock


\bibitem[Wang et~al\mbox{.}(2023)]%
        {ari}
\bibfield{author}{\bibinfo{person}{Jinwen Wang} {et~al\mbox{.}}}
  \bibinfo{year}{2023}\natexlab{}.
\newblock \showarticletitle{{ARI}: Attestation of Real-time Mission Execution
  Integrity}. In \bibinfo{booktitle}{\emph{32nd USENIX Security Symposium}}.
  \bibinfo{pages}{2761--2778}.
\newblock


\bibitem[Wang et~al\mbox{.}(2015)]%
        {bincfi}
\bibfield{author}{\bibinfo{person}{Minghua Wang} {et~al\mbox{.}}}
  \bibinfo{year}{2015}\natexlab{}.
\newblock \showarticletitle{Binary code continent: Finer-grained control flow
  integrity for stripped binaries}. In \bibinfo{booktitle}{\emph{Proceedings of
  the 31st annual computer security applications conference (ACSAC)}}.
  \bibinfo{pages}{331--340}.
\newblock


\bibitem[Wartell et~al\mbox{.}(2012)]%
        {reins_bin_memsafe}
\bibfield{author}{\bibinfo{person}{Richard Wartell} {et~al\mbox{.}}}
  \bibinfo{year}{2012}\natexlab{}.
\newblock \showarticletitle{Securing untrusted code via compiler-agnostic
  binary rewriting}. In \bibinfo{booktitle}{\emph{Proceedings of the 28th
  Annual Computer Security Applications Conference (ACSAC)}}.
  \bibinfo{pages}{299--308}.
\newblock


\bibitem[Wenzl et~al\mbox{.}(2019)]%
        {bin_rewrite_survey}
\bibfield{author}{\bibinfo{person}{Matthias Wenzl} {et~al\mbox{.}}}
  \bibinfo{year}{2019}\natexlab{}.
\newblock \showarticletitle{From hack to elaborate technique—a survey on
  binary rewriting}.
\newblock \bibinfo{journal}{\emph{ACM Computing Surveys (CSUR)}}
  \bibinfo{volume}{52}, \bibinfo{number}{3} (\bibinfo{year}{2019}),
  \bibinfo{pages}{1--37}.
\newblock


\bibitem[Williams-King et~al\mbox{.}(2020)]%
        {egalito}
\bibfield{author}{\bibinfo{person}{David Williams-King} {et~al\mbox{.}}}
  \bibinfo{year}{2020}\natexlab{}.
\newblock \showarticletitle{Egalito: Layout-agnostic binary recompilation}. In
  \bibinfo{booktitle}{\emph{Proceedings of the 25th International Conference on
  Architectural Support for Programming Languages and Operating Systems
  (ASPLOS)}}. \bibinfo{pages}{133--147}.
\newblock


\bibitem[Xu et~al\mbox{.}(2020)]%
        {vulmet}
\bibfield{author}{\bibinfo{person}{Zhengzi Xu} {et~al\mbox{.}}}
  \bibinfo{year}{2020}\natexlab{}.
\newblock \showarticletitle{Automatic hot patch generation for android
  kernels}. In \bibinfo{booktitle}{\emph{29th USENIX Security Symposium}}.
  \bibinfo{pages}{2397--2414}.
\newblock


\bibitem[Yadav and Ganapathy(2023)]%
        {blast}
\bibfield{author}{\bibinfo{person}{Nikita Yadav} {and} \bibinfo{person}{Vinod
  Ganapathy}.} \bibinfo{year}{2023}\natexlab{}.
\newblock \showarticletitle{Whole-Program Control-Flow Path Attestation}. In
  \bibinfo{booktitle}{\emph{Proceedings of the 2023 ACM SIGSAC Conference on
  Computer and Communications Security (CCS)}}. \bibinfo{pages}{2680--2694}.
\newblock


\bibitem[Yagemann et~al\mbox{.}(2021a)]%
        {yagemann2021arcus}
\bibfield{author}{\bibinfo{person}{Carter Yagemann} {et~al\mbox{.}}}
  \bibinfo{year}{2021}\natexlab{a}.
\newblock \showarticletitle{{ARCUS}: symbolic root cause analysis of exploits
  in production systems}. In \bibinfo{booktitle}{\emph{30th USENIX Security
  Symposium}}. \bibinfo{pages}{1989--2006}.
\newblock


\bibitem[Yagemann et~al\mbox{.}(2021b)]%
        {yagemann2021automated}
\bibfield{author}{\bibinfo{person}{Carter Yagemann} {et~al\mbox{.}}}
  \bibinfo{year}{2021}\natexlab{b}.
\newblock \showarticletitle{Automated bug hunting with data-driven symbolic
  root cause analysis}. In \bibinfo{booktitle}{\emph{Proceedings of the 2021
  ACM SIGSAC Conference on Computer and Communications Security (CCS)}}.
  \bibinfo{pages}{320--336}.
\newblock


\bibitem[Zeitouni et~al\mbox{.}(2017)]%
        {atrium}
\bibfield{author}{\bibinfo{person}{Shaza Zeitouni} {et~al\mbox{.}}}
  \bibinfo{year}{2017}\natexlab{}.
\newblock \showarticletitle{{ATRIUM}: Runtime attestation resilient under
  memory attacks}. In \bibinfo{booktitle}{\emph{2017 IEEE/ACM International
  Conference on Computer-Aided Design (ICCAD)}}. IEEE,
  \bibinfo{pages}{384--391}.
\newblock


\bibitem[Zeng et~al\mbox{.}(2022)]%
        {zeng2022palantir}
\bibfield{author}{\bibinfo{person}{Jun Zeng} {et~al\mbox{.}}}
  \bibinfo{year}{2022}\natexlab{}.
\newblock \showarticletitle{Palant{\'\i}r: Optimizing attack provenance with
  hardware-enhanced system observability}. In
  \bibinfo{booktitle}{\emph{Proceedings of the 2022 ACM SIGSAC Conference on
  Computer and Communications Security (CCS)}}. \bibinfo{pages}{3135--3149}.
\newblock


\bibitem[{Zephyr Project}(2016)]%
        {zephyr_project}
\bibfield{author}{\bibinfo{person}{{Zephyr Project}}.}
  \bibinfo{year}{2016}\natexlab{}.
\newblock \bibinfo{title}{{Zephyr Repository}}.
\newblock \bibinfo{howpublished}{\url{https://github.com/zephyrproject-rtos}}.
\newblock
\newblock
\shownote{Accessed: 2025-May-04}.


\bibitem[Zhang et~al\mbox{.}(2013)]%
        {ccfir}
\bibfield{author}{\bibinfo{person}{Chao Zhang} {et~al\mbox{.}}}
  \bibinfo{year}{2013}\natexlab{}.
\newblock \showarticletitle{Practical control flow integrity and randomization
  for binary executables}. In \bibinfo{booktitle}{\emph{IEEE Symposium on
  Security and Privacy (SP)}}. IEEE, \bibinfo{pages}{559--573}.
\newblock


\bibitem[Zhang and Sekar(2015)]%
        {cots_cfi}
\bibfield{author}{\bibinfo{person}{Mingwei Zhang} {and} \bibinfo{person}{R
  Sekar}.} \bibinfo{year}{2015}\natexlab{}.
\newblock \showarticletitle{Control flow and code integrity for COTS binaries:
  An effective defense against real-world ROP attacks}. In
  \bibinfo{booktitle}{\emph{Proceedings of the 31st Annual Computer Security
  Applications Conference (ACSAC)}}. \bibinfo{pages}{91--100}.
\newblock


\bibitem[Zhang et~al\mbox{.}(2017)]%
        {embroidery}
\bibfield{author}{\bibinfo{person}{Xuewen Zhang} {et~al\mbox{.}}}
  \bibinfo{year}{2017}\natexlab{}.
\newblock \showarticletitle{Embroidery: Patching vulnerable binary code of
  fragmentized android devices}. In \bibinfo{booktitle}{\emph{International
  Conference on Software Maintenance and Evolution (ICSME)}}. IEEE,
  \bibinfo{pages}{47--57}.
\newblock


\bibitem[Zhang et~al\mbox{.}(2021)]%
        {recfa}
\bibfield{author}{\bibinfo{person}{Yumei Zhang} {et~al\mbox{.}}}
  \bibinfo{year}{2021}\natexlab{}.
\newblock \showarticletitle{{ReCFA}: resilient control-flow attestation}. In
  \bibinfo{booktitle}{\emph{Annual Computer Security Applications Conference
  (ACSAC)}}. \bibinfo{pages}{311--322}.
\newblock


\bibitem[Zhou et~al\mbox{.}(2022)]%
        {zhou2022ncscope}
\bibfield{author}{\bibinfo{person}{Hao Zhou} {et~al\mbox{.}}}
  \bibinfo{year}{2022}\natexlab{}.
\newblock \showarticletitle{{NCScope}: hardware-assisted analyzer for native
  code in android apps}. In \bibinfo{booktitle}{\emph{Proceedings of the 31st
  ACM SIGSOFT International Symposium on Software Testing and Analysis
  (ISSTA)}}. \bibinfo{pages}{629--641}.
\newblock


\end{thebibliography}

\appendix
\section*{APPENDIX}

\section{Alternative Prover Configurations}\label{appdx:compat}

{
\color{black}
%%% Discussion of how SABRE can handle a Prover that employs other schemes – e.g., memory randomization or optimizations like SpecCFA – atop the typical CFA 
%%% Avenues for future work
Alternative mechanisms such as Address Space Layout Randomization (ASLR)~\cite{aslr} or \CFA optimizations (e.g., SpecCFA~\cite{speccfa}) may be concurrently employed on \prv for increased security/performance. In both cases, \acron remains effective as long as \vrf knows the configuration of the respective mechanism. For example, for \CFA in MCUs (with no MMU), ASLR would be performed before \prv deployment. Therefore, \vrf would have access to the randomized binary.
Since an ASLR bypass would corrupt a pointer to a memory address post-randomization, \acron would not need to perform any additional handling and can pass the randomized binary with the corresponding \textbf{E2} evidence to \acron.
\CFA optimizations on \textbf{E2} evidence, including speculation-based ones such as SpecCFA~\cite{speccfa}, are lossless by design. In it, \vrf replaces expected subpaths with reduced-sized reserved symbols. Since the subpath-to-symbol correspondence is known by \vrf, its ability to analyze received evidence is not lost, regardless of this optimization.
%Thus, \vrf can expand the optimized evidence to obtain the full run-time trace. Therefore with optimization mechanisms, \vrf must always first map the optimized \textbf{E2} into the expanded run-time trace before passing it to \acron.
}

\section{Limitations and Future Work}\label{appdx:limits}

{
\color{black}

%% Discuss open challenges for scaling CFA/run-time auditing to complex software systems and how it relates to SABRE
%%% Avenues for future work
\textbf{Scaling to Complex Systems.} Most current \CFA and run-time auditing targets MCUs~\cite{sok_cfa_cfi}. Conversely, \acron considers MCU binaries (i.e., MSP430 and ARM Cortex-M33). An open problem in this space is scaling \prv RoTs from low-end MCUs to complex systems. 
Assuming \CFA and generation of \textbf{E2} in complex systems is possible, a scalability challenge would come from performing \acron's symbolic data-flow analysis. 
Although \acron does not face typical path explosion risk in symbolic execution due to its use of concrete run-time traces to guide the execution, a unique concern would arise when emulating the active state (memory and registers) during symbolic data flow analysis, which for MCU binaries is already the bottleneck in \acron's Exploit Locator (recall Sec.~\ref{subsec:runtime_eval}). Optimizing \acron's state modeling during symbolic data-flow analysis is an interesting avenue for future work.

\textbf{Adding Vulnerabilities and Attacks.}
%% Discuss that buffer overflPows/user-after-frees are targeted due to their prevalence, and the challenges of addressing other memory vulnerabilities besides control flow hijacking
%%% Discuss scenarios when sabre's patching might not succeed and what happens as a result.
%%% Avenues for future work
This work focuses on control flow hijacking and code-reuse attacks. For this reason, we tailored our framework towards the most prevalent root-cause memory vulnerabilities for these attacks: use-after-free vulnerabilities and buffer overflow-induced out-of-bounds writes~\cite{mitre-cwe-top-25-2024}. Additionally, these memory vulnerabilities are prevalent for launching arbitrary code execution in real-world MCU software (as discussed in Sec.~\ref{subsec:attacks}). However, other memory vulnerabilities (e.g., signedness errors, integer overflows, and double frees) could lead to arbitrary code execution. In these cases, \acron would determine that a generated patch was ineffective and generate a report to provide a starting point for manual inspection (as discussed in Sec.~\ref{subsec:patch_validation}).
Similarly, \acron does not support data-oriented attacks (e.g., direct-data-manipulation or data-oriented-programming~\cite{data_oriented_attacks}) and would fail in the same manner in its attempts to generate a patch or pass \textbf{E2} verification if the attack did not corrupt any control-data. These attacks could be detected through \DFA run-time evidence that includes data inputs (recall Sec.~\ref{sec:cfa}). Future work could extend \acron to detect/patch other memory vulnerabilities and to use data inputs from \DFA to address data-oriented attacks.
}

\end{document}